\documentclass[a4paper,oneside,11pt]{article}
\pdfoutput=1
\usepackage{cprform}
\usepackage{amsmath,amstext,amsfonts,amsbsy,amssymb,amscd,bbm,epsfig,lscape,slashed,color}

\usepackage{epsfig,cite}
\usepackage{amssymb}
\usepackage{subfigure}
\usepackage{hyperref}

\usepackage{tikz}
\usetikzlibrary{arrows,shapes}
\usetikzlibrary{trees}
\usetikzlibrary{matrix,arrows}                          
\usetikzlibrary{positioning}                            
\usetikzlibrary{calc,through}                           
\usetikzlibrary{decorations.pathreplacing}  
\usetikzlibrary{decorations.pathmorphing}       
\usetikzlibrary{decorations.markings}
\tikzset{
    vector/.style={decorate, decoration={snake}, draw},
        provector/.style={decorate, decoration={snake,amplitude=2.5pt}, draw},
        antivector/.style={decorate, decoration={snake,amplitude=-2.5pt}, draw},
    fermion/.style={draw=black, postaction={decorate},
        decoration={markings,mark=at position .55 with {\arrow[draw=black]{>}}}},
    fermionbar/.style={draw=black, postaction={decorate},
        decoration={markings,mark=at position .55 with {\arrow[draw=black]{<}}}},
    fermionnoarrow/.style={draw=black},
    gluon/.style={decorate, draw=black,
        decoration={coil,amplitude=4pt, segment length=5pt}},                           
    scalar/.style={dashed,draw=black, postaction={decorate},
        decoration={markings,mark=at position .55 with {\arrow[draw=black]{>}}}},
    scalarbar/.style={dashed,draw=black, postaction={decorate},
        decoration={markings,mark=at position .55 with {\arrow[draw=black]{<}}}},
    scalarnoarrow/.style={dashed,draw=black},
    electron/.style={draw=black, postaction={decorate},
        decoration={markings,mark=at position .55 with {\arrow[draw=black]{>}}}},
        bigvector/.style={decorate, decoration={snake,amplitude=4pt}, draw},
}

\newcommand{\ba}{\begin{array}}
\newcommand{\ea}{\end{array}}

\newcommand{\req}[1]{Eq.~(\ref{#1})}
\newcommand{\res}[1]{Section~\ref{#1}}
\newcommand{\reapp}[1]{Appendix~\ref{#1}}

\newcommand{\refig}[1]{Fig.~\ref{#1}}
\newcommand{\ret}[1]{Table~\ref{#1}}

\newcommand{\dif}{{\rm d}}

\newcommand{\Dslash}{\relax{\kern+.25em / \kern-.70em D}}

\newcommand{\fm}{{\rm fm}}
\newcommand{\MeV}{{\rm MeV}}
\newcommand{\GeV}{{\rm GeV}}

\newcommand{\Real}{\relax{\mathsf{\Gamma\kern-.35em R}}}
\newcommand{\Int}{\relax{\mathsf{Z\kern-.40em Z}}}

\newcommand{\half}{{\scriptstyle{{1\over 2}}}}

\newcommand{\NC}{N}
\newcommand{\NF}{N_\mathrm{\scriptstyle f}}

\newcommand{\MSbar}{{\overline{\rm MS}}}

\newcommand{\gbar}{\kern1pt\overline{\kern-1pt g\kern-0pt}\kern1pt}
\newcommand{\mbar}{\kern2pt\overline{\kern-1pt m\kern-1pt}\kern1pt}
\newcommand{\obar}[1]{\kern3pt\overline{\kern-2pt #1\kern-0pt}\kern1pt}

\newcommand{\ZS}{Z_{\rm\scriptscriptstyle S}}
\newcommand{\ZA}{Z_{\rm\scriptscriptstyle A}}
\newcommand{\ZV}{Z_{\rm\scriptscriptstyle V}}

\newcommand{\abar}{\kern1pt\overline{\kern-1pt a\kern-0.5pt}\kern1pt}

\newcommand{\cC}{{\cal C}}
\newcommand{\cD}{{\cal D}}

\newcommand{\cH}{{\cal H}}

\newcommand{\cJ}{{\cal J}}

\newcommand{\cL}{{\cal L}}

\newcommand{\cO}{{\cal O}}

\newcommand{\cQ}{{\cal Q}}
\newcommand{\cR}{{\cal R}}

\newcommand{\cV}{{\cal V}}

\newcommand{\cZ}{{\cal Z}}

\newcommand{\vx}{\mathbf{x}}
\newcommand{\vy}{\mathbf{y}}

\newcommand{\vp}{\mathbf{p}}

\newcommand{\vn}{\mathbf{n}}

\newcommand{\runfac}{U}

\begin{document}


\begin{titlepage}


\vspace*{-30truemm}
\begin{flushright}
IFT-UAM/CSIC-14-003\\
FTUAM-14-3\\[10pt]
\end{flushright}
\vspace{15truemm}


\centerline{\Bigrm Exploring the role of the charm quark in the {\Large $\Delta I=1/2$} rule}
\vskip 10 true mm
\centerline{\bigrm  E.~Endress$^a$ and C.~Pena$^{a,b}$}
\vskip 4 true mm
\centerline{\it $^a$ Instituto de F\'{\i}sica Te\'orica UAM/CSIC}
\centerline{\it c/Nicol\'as Cabrera 13-15, Universidad Aut\'onoma de Madrid}
\centerline{\it Cantoblanco E-28049 Madrid, Spain}
\vskip 3 true mm
\centerline{\it $^b$ Departamento de F\'{\i}sica Te\'orica, Universidad Aut\'onoma de Madrid}
\centerline{\it Cantoblanco E-28049 Madrid, Spain}
\vskip 30 true mm


\noindent{\tenbf Abstract:}
{\tenrm
We study the dependence on the charm quark mass of the leading-order low-energy
constants of the $\Delta S=1$ effective Hamiltonian, with the aim of elucidating the role of the charm mass
scale in the $\Delta I=1/2$ rule for $K\to\pi\pi$ decay. To that purpose, finite-volume
Chiral Perturbation Theory predictions are matched to QCD simulations, performed in the
quenched approximation with overlap fermions and $m_u=m_d=m_s$.
Light quark masses range between a few MeV up to around one third of the physical strange mass,
while charm masses range between $m_u$ and a few hundred MeV.
Novel variance reduction techniques are used to obtain a signal for penguin contractions
in correlation functions involving four-fermion operators.
The important role played by the subtractions required to construct renormalised
amplitudes for $m_c \neq m_u$ is discussed in detail.
We find evidence that the moderate enhancement of the $\Delta I=1/2$ amplitude
previously found in the GIM limit $m_c=m_u$
increases only slightly as $m_c$ abandons the light quark regime.
Hints of a stronger enhancement for even higher values of $m_c$ are also found,
but their confirmation requires a better understanding of the subtraction terms.
}
\vspace{10truemm}

\eject
\end{titlepage}

\section{Introduction}
\label{sec:intro}

The quantitative understanding of non-leptonic kaon decays, such as $K\to\pi\pi$,
remains an elusive problem after several decades of study. Thus, no fully solid Standard
Model computation of the value of $\epsilon'/\epsilon$, or of the amplitudes involved
in the famous $\Delta I=1/2$ rule, is available. In this paper we focus on the latter
problem. The decay of a neutral kaon into a pair of pions with total isospin $I$
has an associated transition amplitude
\begin{gather}
T[K\to(\pi\pi)_{I}] = iA_Ie^{i\delta_I}\,,
\end{gather}
where $\delta_I$ is the pion scattering phase shift. Experiment finds that
the amplitude in the $I=0$ channel is significantly larger than the one in
the $I=2$ channel,
\begin{gather}
\frac{|A_0|}{|A_2|} \simeq 22.1\,.
\end{gather}
Early analysis of the $\Delta I=1/2$
problem showed that, if its explanation is to be found in the Standard Model, the bulk
of the enhancement must come from long-distance contributions generated by the
strong interaction~\cite{Gaillard:1974nj,Altarelli:1974exa}.
Reliable determinations of the latter inevitably require a
non-perturbative computation~\cite{Cabibbo:1983xa,Brower:1984ta}.\footnote{An up-to-date
review of kaon decay, including a discussion of the $\Delta I=1/2$ rule,
can be found in~\cite{Cirigliano:2011ny}. See also~\cite{Buras:2014maa} for a discussion
of state-of-the-art attempts to address the phenomenon in the context of large $N$ methods.}

The lattice regularisation of QCD is the only
known approach capable of providing fully first-principles results at
the non-perturbative level. Yet, lattice studies of $K\to\pi\pi$ have to face
significant difficulties:
\begin{itemize}

\item The computation of transition amplitudes
for two-body decays from the Euclidean correlation functions provided by lattice QCD
requires non-trivial kinematical setups~\cite{Maiani:1990ca,Lellouch:2000pv,Lin:2001ek},
which ultimately has a significant impact on the computational cost.

\item The renormalisation of the relevant weak effective Hamiltonian $H_{\rm w}$ is complex. When
the charm quark is not kept as an active degree of freedom the four-quark operators in $H_{\rm w}$
are power-divergent, and non-perturbative subtractions are needed to obtain finite amplitudes.
Furthermore, even when the charm is not integrated out the same is true unless the regularisation
preserves chiral symmetry. Thus, lattice studies with Wilson fermions are poised to deal with
this problem.\footnote{Twisted-mass regularisations with Wilson-like fermions have however been devised
that allow to alleviate or eliminate power divergences~\cite{Pena:2004gb,Frezzotti:2004wz}.}
The use of lattice fermion regularisations with Ginsparg-Wilson fermions~\cite{Ginsparg:1981bj,Kaplan:1992bt,Kaplan:1992sg,Shamir:1993zy,Furman:1994ky,Hasenfratz:1997ft,Hasenfratz:1998jp,Neuberger:1997fp,Luscher:1998pqa,Hernandez:1998et},
that possess an exact chiral symmetry and have been shown to preserve good renormalisation
properties of the operators~\cite{Capitani:2000bm}, is therefore advantageous.
This, however, has again an impact on the computational cost,
since Ginsparg-Wilson fermions are numerically expensive.

\end{itemize}
In recent years, computations that employ so-called domain wall fermions have succeeded
in making significant progress in the study of non-leptonic kaon decays, by computing
amplitudes involving the effective Hamiltonian without a charm quark~\cite{Noaki:2001un,Blum:2001xb,Blum:2011pu,Boyle:2012ys}.

There are several possible sources for the $\Delta I=1/2$ enhancement within the context of strong
interactions. This is ultimately connected to the presence of various scales in the problem:
the charm quark mass scale $m_c \sim 1.3~\GeV$; the intrinsic QCD scale $\Lambda_{\rm QCD} \sim 250~\MeV$;
and the scale $\lesssim 100~\MeV$ of pion final state interactions.
In particular, the role of the charm quark and its associated mass scale
as a possible cause for the $\Delta I=1/2$ enhancement was pointed out long ago~\cite{Shifman:1975tn}.
However, charm effects are not easily apprehended when its contribution to $H_{\rm w}$
is integrated out. This, together with the much simpler renormalisation properties
resulting from the presence of a working GIM mechanism,
constitutes a strong case to keep the charm as an active degree of freedom in the low-energy
treatment of electroweak effects.

In~\cite{Giusti:2004an} a strategy was proposed to disentangle contributions from the various scales, and 
quantify them using numerical simulations. The starting point is the CP-conserving $\Delta S=1$
effective weak Hamiltonian with an active charm quark. One then constructs its counterpart
within the low-energy effective description of QCD provided by Chiral Perturbation
Theory (ChiPT). This is done for two different physical situations: the physical kinematics,
where the charm is heavy and the relevant symmetry for the chiral dynamics is
${\rm SU}(3)_{\rm L} \times {\rm SU}(3)_{\rm R}$; and the unphysical GIM limit $m_c=m_u$,
where the charm is light and the relevant chiral symmetry is
${\rm SU}(4)_{\rm L} \times {\rm SU}(4)_{\rm R}$. In either case, the low-energy
constants (LECs) of the chiral effective Hamiltonian can be determined by matching suitable
correlation functions in ChiPT and QCD.
The use of the effective description, first proposed in~\cite{Bernard:1985wf},
implies dealing with $K\to\pi$ transitions only, which has a double effect:
it avoids the kinematical difficulties posed by the two-body decay, allowing
for smaller volumes (and hence a reduced computational cost); and it neglects
final-state interaction effects, which isolates one of the possible sources for enhancement.
The calculation of the LECs corresponding to $\Delta I=1/2$ and $\Delta I=3/2$
transitions in the GIM limit will expose the effect from intrinsic
QCD scales. The effect of a heavier charm quark can then be studied by monitoring
the behaviour of the amplitudes as $m_c$ increases towards its physical value, exiting
the domain of validity of ChiPT for the charm sector in the process.

Results in the GIM limit were obtained in~\cite{Giusti:2006mh,Hernandez:2008ft}
from quenched QCD simulations with overlap quarks. It was found that in this case
the $I=2$ amplitude is already very close to its physical value, and that a significant
enhancement is already present. The $I=0$ amplitude is however still smaller than
its physical value by roughly a factor of~4. The question is then left whether
increasing $m_c$ towards heavy values provides the bulk of the missing enhancement.
Extending the study to $m_c \gg m_u$ is however non-trivial, because it requires the computation of new
correlation functions --- in the form of so-called ``penguin contractions'' or ``eye diagrams'' ---
notoriously affected by severe signal-to-noise problems. The construction of renormalised
amplitudes for $m_c \neq m_u$ also requires subtractions that eliminate logarithmic
divergences not present in the GIM limit, which adds an extra layer of complication.

In this work we present the first results of an exploration of the effect of a heavier
charm quark on the $\Delta I=1/2$ amplitude, extending the study in~\cite{Giusti:2006mh}.
We will focus on the physics discussion and results; the variance reduction techniques
developed for the computation are described in a companion paper~\cite{techpaper}.
Simulations will still be carried out in the quenched approximation. This is not expected
to have a major impact on the qualitative results of the analysis, and avoids the large
increase of the computational cost that dynamical overlap simulations would imply ---
or, alternatively, the technical and conceptual complications associated to a mixed-action
strategy, in case one would like to use dynamical configurations obtained with a different
fermion regularisation.

The layout of the paper is as follows. In~\res{sec:strat} we summarise the strategy
introduced in~\cite{Giusti:2004an}. In~\res{sec:subt} we discuss the role of subtraction terms,
and how they can be treated. In~\res{sec:latt} we discuss our lattice results for
the relevant QCD correlation functions. In~\res{sec:match} these results are matched
to ChiPT to extract the values of the leading-order LECs. Finally,~\res{sec:concl}
presents our conclusions and outlook. A number of technicalities are discussed in appendices.

\section{Setup and strategy}
\label{sec:strat}

The setup we follow to disentangle the role of the charm quark in the $\Delta I=1/2$ rule
has been laid out in~\cite{Giusti:2004an}. Here we summarise it, and refer
the reader to that paper for a fully detailed discussion of
the various aspects.

\subsection{Effective weak Hamiltonian with an active charm quark}

When the charm quark is kept as an active degree of freedom, and after neglecting
the contribution from top quark loops,\footnote{The top contribution is suppressed by three orders of magnitude relative to the one from the up quark, so that the relation $V_{cs}^*V_{cd}\simeq V_{us}^*V_{ud}$
between CKM matrix elements holds to a good approximation.} the effective Hamiltonian
that describes $K\to\pi\pi$ decays in the Standard Model at scales well below $M_W$ has the form
\begin{gather}
\label{eq:heff}
H_{\rm w}(x) = \frac{g_{\rm w}^2}{4M_W^2}V_{us}^*V_{ud}\sum_n k_n Q_n(x)\,,
\end{gather}
where $g_{\rm w}^2=4\sqrt{2}G_{\rm F} M_W^2$, the sum runs over all the
composite operators $Q_n$ with engineering dimension $d \leq 6$ and
appropriate transformation properties under the relevant symmetries,
and $k_n$ are the corresponding Wilson coefficients.

The relevant global
symmetry group is ${\rm SU}(4)_{\rm L} \times {\rm SU}(4)_{\rm R}$, and
the left-handed character of electroweak interactions demands that operators
are singlets under ${\rm SU}(4)_{\rm R}$. Only two four-quark operators with the
correct flavour content and transformation properties can be constructed, namely
\begin{gather}
Q_1^\pm = J_\mu^{su}J_\mu^{ud} \pm J_\mu^{sd}J_\mu^{uu}~-~[u\leftrightarrow c]\,,
\end{gather}
where $J_\mu$ is the left-handed current
\begin{gather}
J_\mu^{\alpha\beta} = (\bar\psi_\alpha\gamma_\mu P_-\psi_\beta)\,,
\end{gather}
$P_\pm=\half(\mathbf{1}\pm\gamma_5)$, and parentheses around quark bilinears
indicate that they are traced over spin and colour. $Q_1^+,Q_1^-$ transform in
irreducible representations of ${\rm SU}(4)_{\rm L}$ of dimensions 84 and 20, respectively.
The only two other possible operators are quark bilinears, multiplied by factors
involving the quark mass matrix $M$; when the latter is diagonal, $M={\rm diag}(m_u,m_d,m_s,m_c)$,
the two operators are actually identical, and reduce to
\begin{gather}
\begin{split}
Q_2^\pm &= (m_u^2-m_c^2)\left\{m_d(\bar s P_+ d)\,+\,m_s(\bar s P_- d)\right\}\\
&= \half(m_u^2-m_c^2)\left\{(m_d+m_s)(\bar s d) - (m_s-m_d)(\bar s \gamma_5 d)\right\}\,.
\end{split}
\end{gather}
We will keep the $\pm$ superscript in this operator nonetheless, for the sake of notational consistency.
Note that the effective Hamiltonian in~\req{eq:heff} is much simpler than the one
obtained when the charm quark is integrated out --- in that case, $H_{\rm w}$ will
contain ten operators (of which some are redundant). The two main advantages of keeping
an active charm are that the renormalisation properties of composite operators (see below)
are much simpler due to the presence of a working GIM mechanism; and
it is possible to study the dependence of QCD amplitudes on $m_c$ directly.

For the latter purpose, it turns out to be convenient to also have~\req{eq:heff}
rewritten in terms of operators that transform in irreducible representations
of the flavour group ${\rm SU}(3)_{\rm L} \times {\rm SU}(3)_{\rm R}$ spanned
by the light $u,d,s$ quarks. The outcome of this exercise is~\cite{Hernandez:2006kz}
\begin{gather}
\label{eq:heff3}
H_{\rm w} = \frac{g_{\rm w}^2}{4M_W^2}V_{us}^*V_{ud}\left\{
k_1^+ Q_u^+ + \frac{k_1^+}{5}\, R^+ + k_1^- R^-
- k_1^+ Q_c^+ - k_1^- Q_c^- + k_2^+ Q_2^+ + k_2^- Q_2^-
\right\}\,,
\end{gather}
where
\begin{align}
Q_u^+   &= J_\mu^{su}J_\mu^{ud} + J_\mu^{sd}J_\mu^{uu} - 
           \frac{1}{5}\sum_{q=u,d,s}\left\{ J_\mu^{sq}J_\mu^{qd} + J_\mu^{sd}J_\mu^{qq} \right\} \,,\\
R^\pm   &= \sum_{q=u,d,s}\left\{ J_\mu^{sq}J_\mu^{qd} \pm J_\mu^{sd}J_\mu^{qq} \right\} \,,\\
Q_c^\pm &= J_\mu^{sc}J_\mu^{cd} \pm J_\mu^{sd}J_\mu^{cc}\,.
\end{align}
The operator $Q_u^+$ transforms under the 27-plet of ${\rm SU}(3)_{\rm L}$, while all other
operators transform under irreducible representations of dimension 8. Note the trivial identities
$Q_1^+ = Q_u^+ + \frac{1}{5}R^+ - Q_c^+,~Q_1^- = R^- - Q_c^-$.

\subsection{Renormalisation and mixing}

The full weak Hamiltonian is finite, and does not require any renormalisation.
The operators $Q_{1,2}^\pm$, on the other hand, must be renormalised. Assuming
that the regularisation preserves enough of the relevant symmetries (which will
be the case in what follows), the general relation between bare and renormalised
(denoted with a bar) operators is
\begin{gather}
\begin{split}
\bar Q_1^\pm &= Z_{11}^\pm Q_1^\pm + Z_{12}^\pm Q_2^\pm \,,\\
\bar Q_2^\pm &= Z_{21}^\pm Q_1^\pm + Z_{22}^\pm Q_2^\pm \,.
\end{split}
\end{gather}
Since the operator $Q_2^\pm$ only contains products of non-singlet chiral densities times linear combinations of quark masses, it is multiplicatively renormalisable, which allows to choose $Z_{21}^\pm=0$.
Furthermore, as a consequence of the GIM mechanism
the contribution of $Q_2^\pm$ to renormalised operators vanishes when $m_u=m_c$;
this allows to fix $Z_{11}^\pm$ at vanishing quark masses. It is then enough
to fix $Z_{12}^\pm$ such that any remaining divergences are subtracted. Equivalently,
one can rewrite the effective Hamiltonian as
\begin{gather}
\label{eq:hren}
H_{\rm w} = \sum_{\sigma=\pm} k_1^\sigma(\mu) Z_{11}^\sigma(\mu)\left\{Q_1^\sigma\,+\,c^\sigma Q_2^\sigma\right\}\,,
\end{gather}
where
$Q_i^\pm$ are the bare operators,
and impose two subtraction conditions that determine the coefficients $c^\pm$
in such a way that the only remaining divergence in the subtracted operators
$Q_1^\pm\,+\,c^\pm Q_2^\pm$ are eliminated by $Z_{11}^\pm$.
(This is obviously equivalent to fixing $Z_{12}^\pm$). This procedure will be discussed in detail
below. Note that the operator mixing encoded in $c^\pm$ is a radiative effect, so one expects
$c^\pm$ to be naturally of $\cO(\alpha_{\rm s})$, leading to a
suppression of the contribution of $Q_2^\pm$ to physical amplitudes.\footnote{As we will
discuss below, this suppression can be actually argued to be even stronger.}
Note also that the coefficients $c^\pm$ are expected to contain logarithmic
divergences, since the anomalous dimensions of the bare operators $Q_1^\pm$
and $Q_2^\pm$ are different. In a mass independent renormalisation scheme,
one should isolate the values of $c^\pm$ in the chiral limit
and compute them at the same scale at which the overall renormalisation constants
$Z_{11}^\pm$ and the Wilson coefficients $k_1^\pm$ are computed.

Once the operators are renormalised, they have to be combined with Wilson coefficients
into the weak Hamiltonian. Wilson coefficients can be computed from the perturbative anomalous
dimensions, which are known at next-to-leading
order in various dimensional regularisation-based schemes,
as well as in the regularisation-independent (RI) scheme~\cite{Altarelli:1980fi,Buras:1989xd,Ciuchini:1997bw,Buras:2000if}. Correlation functions involving the operators
will be computed on the lattice, and are best non-perturbatively renormalised;
the two schemes of choice to this purpose are RI and Schr\"odinger Functional (SF) schemes.
The main difference between the two options is that the RI procedure allows to renormalise
the operators at scales in the ballpark of few ${\rm GeV}$, while the SF method provides
renormalisation constants at any value of the scale between $\mu\sim\Lambda_{\rm QCD}$ and $\mu\sim M_W$.
The use of RI thus allows to compute the product $k_n(\mu)\bar Q_n(\mu)$ directly, with the
disadvantage that the value of $\mu$ is relatively low and the uncertainty related to the
perturbative truncation in $k_n$ has to be assessed. With SF, on the other hand, a matching
between renormalisation schemes is needed, but it can be performed at high energy scales,
where the convergence of perturbation theory is very good. This will thus be our method of choice.

A convenient way to embody this procedure is to work in a renormalisation group invariant (RGI)
formulation. To that purpose one defines RGI operators and Wilson coefficients as
\begin{gather}
Q^{\rm\scriptscriptstyle RGI} = \runfac(\mu/\Lambda)Q(\mu) = \runfac(\mu/\Lambda)Z(\mu)Q\,,~~~~~~~
k^{\rm\scriptscriptstyle RGI} = \runfac(\mu/\Lambda)^{-1}k(\mu)\,,
\end{gather}
where the RG running factor $\runfac(\mu/\Lambda)$ that connects the renormalised quantity at scale $\mu$
to its RGI counterpart is given by
\begin{gather}
\runfac(\mu/\Lambda) = \left[2b_0\gbar^2(\mu)\right]^{\frac{\gamma_0}{2b_0}}\exp\left\{
-\int_0^{\gbar(\mu)}\dif g\left[\frac{\gamma(g)}{\beta(g)}+\frac{\gamma_0}{b_0 g}\right]
\right\}\,,
\end{gather}
where $\gamma$ and $\beta$ are the anomalous dimension of $Q$ and the RG $\beta$-function
in the scheme of choice, respectively, and $\gamma_0,b_0$ are the leading-order coefficients of their perturbative 
expansions. The use of the SF scheme allows to compute both $Z(\mu)$ and $\runfac(\mu/\Lambda)$
for small values of $\mu/\Lambda$; in the case of the running factor this is achieved by splitting it as
\begin{gather}
\runfac(\mu/\Lambda) = \runfac(M_W/\Lambda)\,\frac{\runfac(\mu/\Lambda)}{\runfac(M_W/\Lambda)}\,,
\end{gather}
where the second factor on the r.h.s. is computed non-perturbatively, and the first one
is computed at next-to-leading order with a small perturbative truncation error
of order $\alpha_s(M_W)^3 \sim \cO(10^{-3})$. The RGI Wilson coefficient
can instead be computed directly as $k^{\rm\scriptscriptstyle RGI} = \runfac(M_W/\Lambda)^{-1}k(M_W)$,
with the same degree of perturbative uncertainty. In view of the construction of the
weak Hamiltonian, it is convenient to define the quantities
\begin{gather}
\cZ_1^\pm \equiv k_1^{\pm;{\rm RGI}}\runfac_1^\pm(\mu/\Lambda)\frac{Z_{11}^\pm(\mu)}{\ZA^2}\,,
\end{gather}
where $\ZA$ is the normalisation factor of the left-handed current
(which will be non-trivial in the lattice regularisation of QCD
that we will introduced later). Note that $\cZ_1^\pm$ is independent by construction
of the renormalisation scale $\mu$.

The running factor $U(\mu/\Lambda)$ has been computed non-perturbatively in~\cite{Guagnelli:2005zc,Dimopoulos:2007ht}
with $\NF=0$ and $\NF=2$ dynamical flavours, respectively. The renormalisation factors
$Z_{11}^\pm(\mu)/\ZA^2$ for the overlap fermion regularisation that we will employ
in this work have been determined in quenched QCD in~\cite{Dimopoulos:2006ma}.

\subsection{Effective low-energy description in Chiral Perturbation Theory}

As discussed in the introduction, a direct computation of $K\to\pi\pi$ amplitudes,
requiring large physical volumes, is beyond the current scope of our work.
We thus resort to computing instead the LECs in the ChiPT counterpart
of the effective weak Hamiltonian, from which the amplitudes can be computed
at some given order in the chiral expansion. Since our main emphasis is to
understand their dependence on $m_c$, we will face two different physical
situations: the strict GIM limit, where all quark masses are light and degenerate;
and the ``physical'' kinematics, where $m_u=m_d=m_s$ are kept light and $m_c \gg m_u$.
In the former case, all four quarks can be treated within ChiPT, while in the latter
only the light flavours enter the effective description;
therefore, two different versions of the chiral effective Hamiltonian will be needed,
with ${\rm SU}(4)$ and ${\rm SU}(3)$ symmetries, respectively.

The construction of the relevant chiral effective weak Hamiltonians has been reviewed in~\cite{Hernandez:2006kz}.
Given a leading-order chiral Lagrangian of the form (either for $U\in{\rm SU}(\NF=4)$ or $U\in{\rm SU}(\NF=3)$)\footnote{Note
that $F$ and $\Sigma$ will of course be different in general depending on the value of $\NF$.}
\begin{gather}
\cL=\frac{F^2}{4}{\rm Tr}\left[(\partial_\mu U)\partial_\mu U^\dagger\right]
-\frac{\Sigma}{2}{\rm Tr}\left[UM^\dagger e^{i\theta/\NF}+MU^\dagger e^{-i\theta/\NF}\right]\,,
\end{gather}
where $M$ is the mass matrix and $\theta$ the vacuum angle,
the leading-order ${\rm SU}(4)$ Hamiltonian reads\footnote{In what follows
the operators $\cQ_2^\pm$, which are the chiral counterparts of $Q_2^\pm$,
will play no role, since ${\rm SU}(4)$ ChiPT will only be used in the limit $m_u=m_c$,
where they drop from $\cH_{\rm w}^{(4)}$. Their explicit form can be found in~\cite{Giusti:2004an}.}
\begin{gather}
\label{eq:hchi4}
\cH_{\rm w}^{(4)} = \frac{g_{\rm w}^2}{4M_W^2}V_{us}^*V_{ud}\sum_{\sigma=\pm}\left\{
g_1^\sigma\cQ_1^\sigma\,+\,g_2^\sigma\cQ_2^\sigma
\right\}\,,
\end{gather}
where $g_{1,2}^\pm$ are LECs,
\begin{gather}
\cQ_1^\pm = \cJ_\mu^{su}\cJ_\mu^{ud} \pm \cJ_\mu^{sd}\cJ_\mu^{uu} \,-\, [u \leftrightarrow c]\,,
\end{gather}
$\cJ_\mu$ is the left-handed chiral current
\begin{gather}
\cJ_\mu = \frac{F^2}{\sqrt{2}}\,U\partial_\mu U^\dagger\,,
\end{gather}
and superscripts indicate matrix components in flavour space.
The ${\rm SU}(3)$ Hamiltonian has instead the form
\begin{gather}
\label{eq:hchi3}
\cH_{\rm w}^{(3)} = \frac{g_{\rm w}^2}{4M_W^2}V_{us}^*V_{ud}\left\{
g_{27}\cQ_{27}\,+\,g_{8}\cQ_{8}\,+\,g'_{8}\cQ'_{8}
\right\}\,,
\end{gather}
where
\begin{align}
\cQ_{27} &= \frac{2}{5}\,\cJ_\mu^{su}\cJ_\mu^{ud} + \frac{3}{5}\, \cJ_\mu^{sd}\cJ_\mu^{uu}\,,\\
\cQ_{8} &= \frac{1}{2}\sum_{q=u,d,s}\cJ_\mu^{sq}\cJ_\mu^{qd}\,,\\
\cQ'_{8} &= m_l\Sigma F^2\left[Ue^{i\theta/\NF}+U^\dagger e^{-i\theta/N_f}\right]^{sd}\,,
\end{align}
where $m_l \equiv m_u=m_d=m_s$. Indeed, in order to avoid
unessential complications related to the soft breaking of the ${\rm SU}(3)$ vector symmetry,
we will always work in the limit of degenerate up, down, and strange masses, which will be
assumed hereafter.

LECs will be determined by matching QCD correlation functions containing the weak
Hamiltonian with ChiPT correlation functions containing its chiral counterpart.
Matching conditions can be imposed separately in different symmetry sectors,
by identifying sets of operators on both sides that transform in the same way
under the relevant chiral symmetry.
In the case of the matching to ${\rm SU}(4)$ ChiPT this is straightforward: $\cQ_{1,2}^\pm$ and $Q_{1,2}^\pm$
have exactly the same transformation properties under ${\rm SU}(4)_{\rm L}$.
In the case of ${\rm SU}(3)$ ChiPT, on the other hand, one finds that $\cQ_{27}$ transforms
in the 27-plet of ${\rm SU}(3)_{\rm L}$, while $\cQ_8$ and $\cQ'_8$ transforms
as octets; since on the QCD side there are one 27-plet and several octet operators,
the matching will be somewhat more involved. Furthermore,
as is well-known, $K\to\pi\pi$ amplitudes
depend on $g_{27}$ and $g_8$ but not on $g'_8$~\cite{Bernard:1985wf,Crewther:1985zt}, rendering the latter
essentially arbitrary; as a matter of fact, the appearance of $g'_8$ reflects
the need for subtractions in QCD amplitudes, as will be discussed in greater detail below.

Note that, since the charm quark is always kept as an active degree of freedom
in QCD, this will imply that the ${\rm SU}(3)$ LECs $g_{27},g_8$ will be functions
of $m_c$. One can actually consider the matching of the chiral Hamiltonians $\cH_{\rm w}^{(4)}$
and $\cH_{\rm w}^{(3)}$ in a regime where $m_c > m_u=m_d=m_s$ but such
that the charm can still be treated within ChiPT, from which point of view
charmed mesons behave as decoupling particles.
This has been studied in~\cite{Hernandez:2004ik}, where explicit expressions for $g_{27}(m_c),g_8(m_c)$
in terms of LO and (unknown) next-to-leading order LECs in ${\rm SU}(4)$ ChiPT are provided.
The leading-order matching reads
\begin{gather}
\label{eq:match_su3su4}
g_{27}(0) = g_1^+\,,~~~~~~~~~~~~~~~
g_8(0) = g_1^-\,+\,\frac{1}{5}\,g_1^+\,.
\end{gather}
On the other hand, one can take the leading-order results for $|A_0|$ and $|A_2|$
in ${\rm SU}(3)$ ChiPT and match them to the experimental values of the amplitudes,
interpreting the result as a phenomenological determination of the LECs at the
physical value $\overline{m}_c$ of the charm quark mass. The result of this
exercise is
\begin{gather}
\label{eq:lecs_pheno}
|g_{27}^{\rm exp}(\overline{m}_c)| \sim 0.50\,,~~~~~~~~~~~~~~~
|g_8^{\rm exp}(\overline{m}_c)| \sim 10.5\,.
\end{gather}

One important ingredient of our setup is that we work both in the
standard, $p$-regime of ChiPT, and in the so-called $\epsilon$-regime~\cite{Gasser:1987ah,Gasser:1987zq}
(see also~\cite{Neuberger:1987zz,Neuberger:1987fd}). Here $p$-regime means
working in large volumes measured in terms of the pion Compton wavelength,
i.e. $m_{\pi}L \gg 1$ if a four-dimensional box of dimensions $L^3 \times T$ is considered;
$\epsilon$-regime means keeping a large volume (i.e. the implicit $F_\pi L \gg 1$ prerrequisite
for the chiral expansion to work is fulfilled) but working at very small quark
masses, such that the ``pion'' Compton wavelength is of the order of $L$ --- or, more
precisely, $m\Sigma V \lesssim 1$, where $m$ is the light quark mass, $\Sigma$ is the
chiral condensate, and $V$ is the four-dimensional volume. Furthermore, one should keep
$T \sim L$, since at $T/L \gg 1$ a different kinematical region --- the $\delta$-regime~\cite{Leutwyler:1987ak} --- arises.
The main advantage of considering the $\epsilon$-regime instead of the physical $p$-regime
is that mass effects are suppressed in the former, and the chiral expansion is
rearranged such that less operators appear at any given order in the expansion
with respect to the $p$-regime~\cite{Hansen:1990yg}. This allows for potentially cleaner determinations
of the leading-order LECs --- especially so in the case of effective Hamiltonians for
non-leptonic meson decay, which display a large number of new terms at NLO in the chiral
expansion~\cite{Kambor:1989tz}. On the other hand, finite-volume effects are obviously
large in the $\epsilon$-regime, being typically polynomial and not exponentially
suppressed as in the $p$-regime. Finally, out of technical convenience
correlation functions in the $\epsilon$-regime are computed at a fixed value of the topological charge.

It can be shown~\cite{Gasser:1987ah} that LECs are universal, in the sense that the same
values are obtained when ChiPT is matched to QCD in either kinematical regime.
Since the systematic uncertainties induced by the truncation of the chiral
expansion are however different in each case, being able to perform consistent
matching in both regimes implies a much higher degree of control on the final results.
In particular, the ChiPT correlation functions involved
in the matching for leading-order LECs in the chiral effective Hamiltonian
will not depend on extra LECs up to NNLO corrections --- NLO contributions
are purely finite-volume effects, which are exactly calculable.
Note that on the QCD side, the need of having non-perturbative results at very low quark
masses and for a well-defined value of the topological charge in order to work
in the $\epsilon$-regime implies that
lattice regularisations with exact chiral symmetry are strongly preferred.

One final comment concerns the use of quenched Chiral Perturbation Theory (qChiPT)
to describe quenched QCD data. As is well-known, qChiPT displays
unphysical artifacts; in particular, in the context of $K\to\pi\pi$ transitions
Golterman-Pallante ambiguities make the matching of QCD to ${\rm SU}(3)$
qChiPT ill-defined~\cite{Golterman:2001qj,Golterman:2002us}. This is however
not the case for ${\rm SU}(4)$, where the ratios of correlation functions
we will deal with (see below) present no ambiguities in the quenched approximation, as
discussed in~\cite{Giusti:2004an,Hernandez:2006kz}. Quenched results
are not worked out explicitly in~\cite{Hernandez:2006kz} for ${\rm SU}(3)$
ChiPT. As can be seen in the formulae gathered in~\reapp{app:chipt}, while
the $\epsilon$-regime formulae are essentially insensitive to quenching, the
NLO prediction $p$-regime predictions for the relevant correlation functions
in the octet channel displays $1/N_{\rm f}$ factors,
that signal the need to take into account non-decoupled singlet contributions
to repeat the computation in the quenched case. Here we will take the
unquenched formulae as an operational description, and perform fits with various
values of $N_{\rm f}$ (and hence different coefficients in the chiral logs)
to check the dependence of the LECs on the value of $N_{\rm f}$, and adscribe
a systematic uncertainty to fit results (see~\res{sec:match} for details).

\subsection{Matching ChiPT to QCD}

\subsubsection{$m_c=m_l$}

When all quarks are light and degenerate the effective low-energy description of
$\Delta S=1$ processes is given by~\req{eq:hchi4}. Contributions from $Q_2^\pm$ (in QCD)
and $\cQ_2^\pm$ (in ChiPT) drop because they are proportional to $m_u-m_c$;
one is thus left with the problem of determining the LECs $g_1^\pm$. As explained
above, the correspondence between QCD and ChiPT operators in this case is straightforward.
The matching can be easily performed using three-point functions of
the operators in the effective Hamiltonian with quark bilinears such that
flavour indices are saturated. A technically convenient choice for the latter
is to employ left-handed currents, leading to the correlation functions
\begin{align}
C_i^\pm(x_0,y_0) &= \int\dif^3x\int\dif^3y\,\langle J_0^{du}(x) \, Q_i^\pm(0) \, J_0^{us}(y)\rangle\,,\\
C(x_0) &= \int\dif^3x\,\langle J_0^{\alpha\beta}(x) \, J_0^{\beta\alpha}(0)\rangle\,,
\end{align}
where $\alpha,\beta$ are distinct light flavour indices (not summed over).
The ratios
\begin{gather}
\label{eq:qcd_rat}
R_i^\pm(x_0,y_0) = \frac{C_i^\pm(x_0,y_0)}{C(x_0)C(y_0)}\,,
\end{gather}
will then be proportional to the matrix elements $\langle\pi|Q_1^\pm|K\rangle$
(with mass-degenerate kaon and pion) when $x_0\to +\infty,y_0\to -\infty$.
The equivalent ChiPT quantities are
\begin{align}
\cC(x_0) &= \int\dif^3x\,\langle \cJ_0^{ud}(x) \, \cJ_0^{du}(0)\rangle_{{\rm SU}(4)}\,,\\
\cC_i^\pm(x_0,y_0) &= \int\dif^3x\int\dif^3y\,\langle \cJ_0^{du}(x) \, \cQ_i^\pm(0) \, \cJ_0^{us}(y)\rangle_{{\rm SU}(4)}\,,\\
\cR_i^\pm(x_0,y_0) &= \frac{\cC_i^\pm(x_0,y_0)}{\cC(x_0)\cC(y_0)}\,,
\end{align}
where the notation $\langle\rangle_{{\rm SU}(4)}$ emphasises the use of the appropriate
effective theory. The LECs in the chiral weak Hamiltonian can then be readily extracted
from the matching condition
\begin{gather}
\cZ_1^\pm R_1^\pm(x_0,y_0) = g_1^\pm \cR_1^\pm(x_0,y_0)\,.
\end{gather}
Formulae for ChiPT quantities are given in~\reapp{app:chipt}.

\subsubsection{$m_c \gg m_l$}

A similar strategy to the one just described can be pursued to match QCD with $m_c \gg m_l$
to ${\rm SU}(3)$ ChiPT. One first defines new three-point functions in both QCD
\begin{gather}
C_u^+(x_0,y_0) = \int\dif^3x\int\dif^3y\,\langle J_0^{du}(x) \, Q_u^+(0) \, J_0^{us}(y)\rangle \,,
\end{gather}
and ChiPT
\begin{align}
\cC_{27}(x_0,y_0) &= \int\dif^3x\int\dif^3y\,\langle \cJ_0^{du}(x) \, \cQ_{27}(0) \, \cJ_0^{us}(y)\rangle_{{\rm SU}(3)}\,,\\
\cC_{8}(x_0,y_0) &= \int\dif^3x\int\dif^3y\,\langle \cJ_0^{du}(x) \, \cQ_8(0) \, \cJ_0^{us}(y)\rangle_{{\rm SU}(3)}\,,\\
\cC'_{8}(x_0,y_0) &= \int\dif^3x\int\dif^3y\,\langle \cJ_0^{du}(x) \, \cQ'_8(0) \, \cJ_0^{us}(y)\rangle_{{\rm SU}(3)}\,,
\end{align}
and the corresponding ratios $R_u^+,\cR_{27},\cR_8,\cR'_8$
by dividing them with products of current two-point functions.
Next one can impose matching conditions in both the 27-plet and octet channels,
\begin{align}
\label{eq:match27}
R_{27}(x_0,y_0) &= g_{27}\cR_{27}(x_0,y_0) \,,\\
\label{eq:match8}
R_8(x_0,y_0) &= g_8\cR_8(x_0,y_0) + g'_8\cR'_8(x_0,y_0)\,,
\end{align}
where
\begin{align}
R_{27} &= \cZ_1^+ R_u^+ \,,\\
R_8 &= \cZ_1^+ \left[R_1^+ - R_u^+ + c^+ R_2^+\right] +
\cZ_1^- \left[R_1^- + c^- R_2^-\right]\,.
\end{align}
Note that there is no contribution from the pure-octet
correlator $R_2^+$ in the 27-plet channel.

It has to be stressed that the matching conditions in Eqs.~(\ref{eq:match27},\ref{eq:match8})
immediately imply that the LECs acquire a dependence on $m_c$.
Furthermore, the matching condition~\req{eq:match8} provides, in principle, only
a linear combination of the two octet LECs; in particular, it does not
directly allow to disentangle the physical ChiPT octet contribution
with $g_8$ from the unphysical one with $g'_8$. As will shown below,
however, typical conditions to determine the subtraction coefficients $c^\pm$
required to construct renormalised QCD amplitudes simultaneously fix
the value of $g'_8$, which is then no longer an unknown. \req{eq:match8}
does then allow to determine $g_8$ unambiguously.
Formulae for ChiPT quantities are again provided in~\reapp{app:chipt}.

\subsection{Results in the GIM limit and scope of the present work}

The ${\rm SU}(4)$ LECs $g_1^\pm$ were determined in~\cite{Giusti:2006mh}
by computing the renormalised ratios of correlation functions
$\cZ_1^\pm R_1^\pm$ in lattice QCD in the quenched approximation
at fixed volume and lattice spacing and keeping $m_c=m_l$.
Computations were performed at four $p$-regime and one $\epsilon$-regime
values of $m_l$; renormalisation factors were separately determined in~\cite{Dimopoulos:2006ma}.
The results were found
\begin{gather}
g_1^+ = 0.51(9)\,,~~~~~~~~~
g_1^- = 2.6(5)\,,
\end{gather}
leading via~\req{eq:match_su3su4} to
\begin{gather}
g_{27}(0) = 0.51(9)\,,~~~~~~~~~
g_8(0) =  2.7(5)\,,
\end{gather}
that can be compared with the phenomenological expectation in~\req{eq:lecs_pheno}.
It can then be concluded that
\begin{enumerate}

\item The approximations involved in the above
computation provide the correct value for the $\Delta I=3/2$
amplitudes parametrised by $g_{27}$ (which are indeed expected to
have little sensitivity to the value of $m_c$).

\item Pure low-energy QCD effects, combined with
the well-known short-distance contribution given by the ratio of
Wilson coefficients $k_1^-/k_1^+$, are responsible
for a significant enhancement of the decay amplitude in
the $\Delta I=1/2$ channel. The latter is however still a factor
$\sim 4$ smaller than the phenomenological value.

\end{enumerate}
Therefore, barring (unlikely) large cutoff effects in the $m_c=m_l$
lattice QCD computation, as well as the possibility of
large quenching artifacts, an explanation of the $\Delta I=1/2$ rule
that is purely based on Standard Model physics requires either
a significant increase in $g_8(m_c)$ when $m_c \gg m_l$;
a strong effect due to pion rescattering in physical $K\to\pi\pi$
decays; or a combination of the two. The aim of the present work
is to explore the dependence of $g_8$ on $m_c$,
by extending the study of~\cite{Dimopoulos:2006ma} to the case $m_c \neq m_l$.
As we will discuss, a major technical challenge for this is the computation
of the new contributions to amplitudes involving the four-fermion operators $Q_1^\pm$
that arise outside the $m_c=m_u$ limit.\footnote{The effect of taking $m_c>m_u$,
for values of $m_c$ that are still light enough to
fit within the effective low-energy description provided by ChiPT, has been studied
in~\cite{Hernandez:2006kz}, by analysing how charm decoupling effects are reabsorbed
in ${\rm SU}(3)$ LECs. This yields a logarithmic enhancement of the $\Delta I=1/2$
amplitude, although lack of knowledge about the corrections coming from NLO terms
in the chiral expansion prevents quantitative statements.}

\section{The role of the subtraction term}
\label{sec:subt}

As discussed above, outside the GIM limit $m_u=m_c$, and for our kinematics
$m_u=m_d=m_s=m_l$, the renormalised matrix elements
$\langle\pi(\vp=\mathbf{0})|\bar Q_1^\pm|K(\vp=\mathbf{0})\rangle$
are a linear combination of the bare $\langle\pi| Q_1^\pm|K\rangle$
matrix elements and the subtraction term $\langle\pi| Q_2^\pm|K\rangle$,
cf.~\req{eq:hren}. In this section we will discuss the contribution of
the subtraction term, as well as two possible procedures to determine
the subtraction coefficients $c^\pm$: fixing $c^\pm$ by prescribing arbitrary
values for the unphysical renormalised amplitudes $\langle 0|\bar Q_1^\pm|K\rangle$; and
a variant of this method that involves two-point functions
of $\bar Q_1^\pm$ in the $\epsilon$-regime.
We will also discuss the behaviour of the subtraction coefficients in perturbation theory.

\subsection{Matrix elements of $Q_2^\pm$}

It is first of all interesting to note that the properties of
amplitudes involving $Q_2^\pm$ are considerably simplified if,
as will be the case in what follows, one is only interested
in matrix elements of the effective weak Hamiltonian with no
momentum transfer between the initial and final state.
Using chiral Ward-Takahashi identities, the contribution from the
operators $(\bar s P_\pm d)$ contained in $Q_2^\pm$ to any
amplitude can be rewritten as\footnote{For the purpose of this
argument, we will assume for the moment that all quantities are renormalised.
Comments on the role of renormalisation will be provided later.}
\begin{gather}
\label{eq:q2_cur}
\langle f|(\bar s P_\pm d)|i\rangle =
\frac{\langle f|\partial_\mu(\bar s \gamma_\mu d)|i\rangle}{m_s-m_d}\,\pm\,
\frac{\langle f|\partial_\mu(\bar s \gamma_\mu\gamma_5 d)|i\rangle}{m_s+m_d}\,.
\end{gather}
When $m_s \neq m_d$, this immediately implies that the matrix element is proportional
to the four-momentum transfer, and vanishes if the latter is zero.\footnote{As a matter of fact,
a trivial extension of this argument implies that the subtraction term does not contribute
to physical $K\to\pi\pi$ decay amplitudes, since in that case one has the physical $m_s\neq m_d$ kinematics
and momentum is conserved.}
When $m_s=m_d$, on the other hand, the first term on the r.h.s. has vanishing numerator and denominator,
and the quark mass dependence of $\langle f|\partial_\mu(\bar s \gamma_\mu d)|i\rangle$
has to be studied in order to find the value of the ratio in the limit $m_s\to m_d$.

In physical, $p$-regime kinematics, and for large Euclidean time separations between the operators,
the QCD three-point functions involved in the matching to ChiPT are proportional
to the transition amplitude $\langle\pi^+|H_{\rm w}|K^+\rangle$.
Taking $|i\rangle = |K^+(p)\rangle,|f\rangle = |\pi^+(k)\rangle$
in~\req{eq:q2_cur}, the contribution from the axial current term vanishes due
to parity conservation, and
the standard parametrisation of meson-meson matrix elements
of the vector current in terms of vector ($f_+$) and scalar ($f_0$) form factors leads to
\begin{gather}
\begin{split}
\langle\pi^+(k)|(\bar s d)|K^+(p)\rangle &=
\frac{\langle \pi^+(k)|\partial_\mu(\bar s \gamma_\mu d)|K^+(p)\rangle}{m_s-m_d}\\
&= \frac{q_\mu\left[(p+k-\Delta)_\mu f_+(q^2) + \Delta_\mu f_0(q^2)\right]}{m_s-m_d}\,,
\end{split}
\end{gather}
where $q=p-k$, $\Delta_\mu=(m_K^2-m_\pi^2)q_\mu/q^2$, and the
normalisation convention $f_+(0)=f_0(0)$ applies. If the external states are on-shell,
the above expression reduces to
\begin{gather}
\langle\pi^+(k)|(\bar s d)|K^+(p)\rangle =
\frac{m_K^2-m_\pi^2}{m_s-m_d}\,f_0(q^2)\,,
\end{gather}
which does not vanish for $m_s \neq m_d$ (in which case the momentum transfer is indeed
non-zero). If now we take our preferred kinematics $m_u=m_d=m_s$
we will have $m_K=m_\pi$, and the momentum transfer vanishes; but
the matrix element is still non-zero, since the ratio $(m_K^2-m_\pi^2)/(m_s-m_d)$
is finite (and proportional to the chiral condensate), and $f_0(0)=f_+(0)=1$ by
the Ademollo-Gatto theorem~\cite{Ademollo:1964sr}. Thus, the renormalisation of $K\to\pi$ amplitudes
still requires a subtraction for mass-degenerate kaon and pion at rest.

Since the relevant matrix elements are entirely determined by the ratio
$(m_K^2-m_\pi^2)/(m_s-m_d)$, one can actually use ChiPT to obtain a
precise prediction for the value of the subtracted matrix element,
\begin{gather}
\langle\pi^+|Q_2^\pm|K^+\rangle = \half(m_u^2-m_c^2)(m_s+m_d)\langle\pi^+|(\bar s d)|K^+\rangle\,.
\end{gather}
In particular, at leading order and with $m_u=m_d=m_s=m_l$ one has
\begin{gather}
\label{eq:q2_chipt}
\langle\pi^+|Q_2^\pm|K^+\rangle \approx \half(m_l^2-m_c^2)m_{\rm PS}^2\,,
\end{gather}
where $m_{\rm PS}$ is the mass of the pseudoscalar light octet mesons.
A discussion of the NLO ChiPT corrections to \req{eq:q2_chipt} is provided in~\reapp{app:chipt}.

A final comment concerning renormalisation is in order.
As mentioned above, the argument employed to arrive at \req{eq:q2_chipt}
assumes that renormalised quantities are used throughout.
In order to make contact with bare lattice quantities, it will be necessary
to take into account relative (re)normalisation factors. For instance,
the result in \req{eq:q2_chipt} will hold for either the bare or renormalised
$K^+\to\pi^+$ amplitude mediated by $Q_2^\pm$, depending on whether the quark
masses in the factor $(m_l^2-m_c^2)$ are bare or renormalised.
In practice, rather than in the amplitude itself we will be interested
in the ratio (to which the quantity $R_2^\pm$ introduced in~\req{eq:qcd_rat}
will tend for large Euclidean time separations)
\begin{gather}
\label{eq:chipt_r2}
\frac{\langle\pi^+|Q_2^\pm|K^+\rangle}{F_{\rm PS}^2 m_{\rm PS}^2} \approx
\frac{m_l^2-m_c^2}{2F^2}\,,
\end{gather}
where $F_{\rm PS}$ is the decay constant of octet pseudoscalar mesons,
and LO ChiPT has again been employed to get to the r.h.s. of the expression.
The factor required to renormalise this ratio is $(\ZS\ZA)^2$, where $\ZS,\ZA$
are the (re)normalisation factors of the non-singlet scalar density and axial currents,
respectively.\footnote{Recall that even if chiral symmetry is exactly preserved
on the lattice by using Neuberger-Dirac fermions, local currents still require
a non-trivial normalisation.} If the ratio on the l.h.s. is the bare one,
and the quark masses on the r.h.s. are also bare, then the relative factor
is given by $\ZA^2$.

Natural prescriptions to fix the subtraction coefficients $c^\pm$
will result in the latter being mass-independent (possibly up
to small corrections, which will depend on the precise procedure to fix them).
Since, on the other hand, we have seen that matrix elements of $Q_2^\pm$
are proportional to $(m_l^2-m_c^2)$, it then follows that for $m_c\gg m_l$ and fixed
$m_l$ the contribution of $Q_2^\pm$ to any amplitude will be,
to good approximation,
proportional to $c^\pm\, m_c^2$. Thus, an interesting question, directly related
to understanding the role of the charm quark in the $\Delta I=1/2$ enhancement,
is whether bare amplitudes involving $Q_1^\pm$ exhibit a similar behaviour;
and whether, if that is the case, some measure of cancellation of this
strong $m_c$ dependence occurs.\footnote{Recall that if the charm had not been
kept as an active degree of freedom in the effective Hamiltonian, the
mixing with dimension-three operators would involve power divergences
that make up for the missing GIM factors; in that case
bare matrix elements of four-fermion operators
contain UV divergences $\propto a^{-2}$, that are cancelled
against the subtractions in physical amplitudes.}

\subsection{Determination of subtraction coefficients}

\subsubsection{Kaon-to-vacuum amplitudes}

A simple way of fixing subtraction coefficients, first proposed in~\cite{Bernard:1985wf},
is to exploit the fact that meson-to-vacuum amplitudes mediated by the effective
weak Hamiltonian do not contribute to any physical process; one can therefore
set them to arbitrary values. The simplest possibility is to impose that
renormalised kaon-to-vacuum amplitudes for $Q_1^\pm + c^\pm Q_2^\pm$ vanish,
\begin{gather}
\label{eq:subt_k0}
\langle 0 | Q_1^\pm + c^\pm Q_2^\pm | K^0\rangle = 0\,.
\end{gather}
The bare amplitudes can be extracted from the QCD two-point functions
\begin{gather}
\label{eq:subt_2p}
D_{1,2}^\pm(x_0) = \int\dif^3x\,\langle Q_{1,2}^\pm(0) \, J_0^{ds}(x)\rangle\,,
\end{gather}
which for large values of $|x_0|$ become proportional
to $\langle 0|Q_{1,2}^\pm| K^0\rangle e^{-m_K|x_0|}$
(up to finite-volume effects). On the other hand, when the kaon-to-vacuum
amplitude is computed in ChiPT one has~\cite{Crewther:1985zt}
\begin{gather}
\label{eq:k0_chipt}
\langle 0|H_{\rm w}| K^0\rangle \propto g'_8[(m_s^2-m_d^2)~+~{\rm higher~orders}]\,,
\end{gather}
which means that fixing the value of the amplitude is equivalent to setting
the value of the unphysical LEC $g'_8$. In particular, \req{eq:subt_k0} implies
$g'_8=0$.

When the explicit form of $Q_2^\pm$ is substituted in~\req{eq:k0_chipt}, it becomes
a linear equation in $c^\pm$ that has the solutions
\begin{gather}
\label{eq:subt_k0_2}
c^\pm = \frac{2}{(m_u^2-m_c^2)(m_s-m_d)}\,\frac{\langle 0|Q_1^\pm| K^0\rangle}{\langle 0|\bar s \gamma_5 d| K^0\rangle}\,,
\end{gather}
where we have used that parity conservation ensures that only the pseudoscalar density part
of $Q_2^\pm$ contributes to the transition. Since $c^\pm$ do not depend on quark masses
by construction, one should ideally compute the ratio of correlation functions at various
values of the quark masses and extrapolate to the chiral limit; in practice, if computations
are carried out at finite quark mass one expects some residual mass dependence.
\req{eq:subt_k0_2}, however, makes a crucial practical shortcoming of this procedure
in our context apparent: when $m_s=m_d$ both the numerator and the denominator vanish,
while leaving a finite limit --- cf.~\req{eq:k0_chipt}, which also (and consistently)
implies that $g'_8$ is not fixed in this case.

One variant of the method that can be applied at $m_s=m_d$ involves matrix elements
with external scalar states, that become the dominant contributions to $D_1^\pm$ in that limit;
denoting by $|{\rm S}\rangle$ the lightest scalar state with one unit of strangeness,
one could impose the condition
\begin{gather}
\label{eq:subt_k0_s}
\langle 0 | Q_1^\pm + c^\pm Q_2^\pm | {\rm S}\rangle = 0\,,
\end{gather}
or, equivalently,
\begin{gather}
\label{eq:subt_k0_2_s}
c^\pm = \frac{2}{(m_u^2-m_c^2)(m_s+m_d)}\,\frac{\langle 0|Q_1^\pm| {\rm S}\rangle}{\langle 0|\bar s d| {\rm S}\rangle}\,.
\end{gather}
Note that these matrix elements are contained in the two-point functions of~\req{eq:subt_2p},
since the left-handed current contains a parity-even component.
In our simulations, the most likely candidate for $|{\rm S}\rangle$ will be a state
containing two pseudoscalar mesons --- a $|K\pi\rangle$ state, given the flavour
assignments. Again, at leading order in the effective description the $|K\pi\rangle\to|0\rangle$
amplitudes receive contributions from $\cQ_8'$ only, and setting the subtraction
condition~\req{eq:subt_k0_s} is equivalent to setting $g'_8=0$, as before. On the other
hand, it can be expected that the determination of these matrix elements from
lattice QCD will be significantly more difficult than in the case where only single meson
states are involved.

\subsubsection{Two-point functions in the $\epsilon$-regime}

A variant of the above procedure consists of computing the correlation functions $D^\pm_{1,2}$ with
$\epsilon$-regime kinematics for the light quarks, as proposed in~\cite{Hernandez:2006kz}.
In that case the computation is carried out at fixed
value of the topological charge $\nu$, and parity is not preserved; as a result,
for a given value of $\nu$ the contribution to $D^\pm_{1;\nu}$ from the pseudoscalar channel
does not vanish at $m_s=m_d$ as in the $p$-regime, avoiding the shortcomings of the method
based on $K^0\to$~vacuum matrix elements.

The two-point functions $D^\pm_{1,2}$ can then be split into ${\rm SU}(3)$ 27-plet and octet contributions
in the same way as was done above for three-point functions,
and matched to the corresponding NLO ChiPT prediction for
\begin{align}
\cD_{27;\nu}(x_0) &= \int\dif^3x\,\langle \cQ_{27}(0)\,\cJ_0^{ds}(x)\rangle_{{\rm SU}(3);\nu}\,,\\
\cD_{8;\nu}(x_0) &= \int\dif^3x\,\langle \cQ_8(0)\,\cJ_0^{ds}(x) \rangle_{{\rm SU}(3);\nu}\,,\\
\cD'_{8;\nu}(x_0) &= \int\dif^3x\,\langle \cQ'_8(0)\,\cJ_0^{ds}(x) \rangle_{{\rm SU}(3);\nu}\,.
\end{align}
In particular, $\cD_{8;\nu}$ vanishes up to NNLO corrections, while $\cD'_{8;\nu}$ does not.
(The 27-plet contribution vanishes identically in both QCD and ChiPT for chiral symmetry reasons.)
The octet contribution is thus given by $\cD'_{8;\nu}$ only, and one has the matching condition
\begin{gather}
\begin{split}
D_{8;\nu}(x_0) &= 
\cZ_1^+ \left[D_{1;\nu}^+ + c^+ D_{2;\nu}^+\right] +
\cZ_1^- \left[D_{1;\nu}^- + c^- D_{2;\nu}^-\right] \\
&= 2g'_8(m_c) \cD'_{8;\nu}(x_0)\,.
\end{split}
\end{gather}
The condition for different values of $\nu$ is not independent, since
the only dependence of $\cD'_{8;\nu}$ on topology is a trivial overall factor~\cite{Hernandez:2006kz}.
As before, the value of $g'_8$ can be set arbitrarily (e.g. to zero);
since, furthermore, this has to hold for all values of the renormalisation scale,
and either operator has different anomalous dimension, the consistency of the
condition then requires that each term vanishes separately, viz.
\begin{gather}
c^\pm = -\,\frac{D_{1;\nu}^\pm}{D_{2;\nu}^\pm}\,,
\end{gather}
which results in a similar subtraction condition to~\req{eq:subt_k0_2}.
In the overlap lattice computation, this expression will be expected to hold
sufficiently far away from operator insertions.

\subsubsection{One-loop analysis}

Alternative to the hadronic conditions to determine subtraction coefficients discussed above,
it is also possible to conduct a perturbative study of the subtraction terms. Note that
having kept the charm quark as an active degree of freedom implies that only logarithmic
divergences appear in renormalisation; as mentioned earlier, this is one of the main
advantages with respect to the
setup where the charm is integrated out, which leads to power divergences whose study is
outside the realm of perturbation theory.
While a full determination of the perturbative value of subtraction coefficients
in a lattice regularisation with Neuberger-Dirac fermions is beyond the scope of this work,
it is already interesting to conduct a one-loop analysis in the continuum.
To our knowledge, such an analysis is not available in the literature.

In order to study the subtraction of the operators $Q_2^\pm$ involved in the
construction of renormalised operators $\bar Q_1^\pm$ in the continuum,
we will impose subtraction conditions of the form
\begin{gather}
\label{eq:risubt}
{\rm tr}\langle
s(p)\,\bar Q_1^\pm\,\bar d(p)
\rangle_{\rm amp} = 0\,,
\end{gather}
where the trace is taken over colour and spin indices, the notation
$\langle\rangle_{\rm amp}$ stands for the amputated correlation
function obtained by multiplying times the inverse quark propagators
running on external legs, and the connection between spacetime and
momentum-space correlation functions is given by
\begin{gather}
\int\dif^4x\,\dif^4y\, e^{ip\cdot(x-y)}
\langle s(x)\,\bar Q_1^\pm(0)\,\bar d(y)\rangle\,.
\end{gather}
The RI-like condition in \req{eq:risubt} is similar to e.g. the one introduced in~\cite{Blum:2001xb}
to determine subtraction coefficients of bilinear operators in the $\Delta S=1$
Hamiltonian with the charm quark integrated out. Furthermore, it is an obvious
perturbative equivalent to hadronic subtraction conditions such as
$\langle 0|\bar Q_1^\pm | K^0\rangle=0$.

A one-loop analysis of \req{eq:risubt} in continuum perturbation theory
is provided in \reapp{app:pert}. The perturbative
computation finds the correct $(m_u^2-m_c^2)(m_s+m_d)$ dependence of the subtraction
term,\footnote{Note that the correlation function in \req{eq:risubt} receives contributions
from the parity-even channel only.} and provides logarithmically divergent values of
$c^\pm$. This is consistent with the misaligned logarithmic divergences in the bare operators
$Q_1^\pm$ and $Q_2^\pm$ that the subtraction coefficients have to account for. Loop integrals
are found to provide factors of $(4\pi)$ such that the one-loop coefficients are of the form
\begin{gather}
c^\pm = \frac{\alpha_{\rm s}}{4\pi}\,\frac{1}{(4\pi)^2}\,\times\,\cO(1)\,.
\end{gather}
(Note that the coefficients can in principle have either sign.)
It is also found that in natural kinematical setups there are no large logs. Taking this
as input, a conservative estimate of the size of subtraction coefficients
is that they are approximately zero, with a systematic uncertainty set to $\alpha_{\rm s}/(4\pi)$;
this is good enough
for the level of precision we will attain in the determination of physical amplitudes
within our explored range in charm masses.


\section{Computation of correlation functions in Lattice QCD}
\label{sec:latt}

\subsection{Regularisation and simulation details}

We simulate lattice QCD using the Wilson plaquette action for the gauge fields,
while quark fields are regularised using a Neuberger-Dirac operator~\cite{Neuberger:1997fp,Neuberger:1998wv}.
The latter satisfies a Ginsparg-Wilson relation of the form
\begin{gather}
\gamma_5 D_{\rm N} + D_{\rm N} \gamma_5 = \abar D_{\rm N}\gamma_5 D_{\rm N}\,,
\end{gather}
where $\abar = a/(1+s)$ and $s$ is a parameter that can be tuned to optimise
the locality properties of the operator.
The techniques we use for the construction, inversion, and spectral studies of $D_{\rm N}$
are discussed in~\cite{Giusti:2002sm}; in our simulations we will always employ $s=0.4$~\cite{Hernandez:1998et}.

The fermion lattice action
\begin{gather}
S_{\rm F} = a^4\sum_x \left\{\bar\psi D_{\rm N}\psi + m\bar\psi\tilde\psi\right\}(x)\,,~~~~~
\tilde\psi=\left(\mathbf{1}-\tfrac{\abar}{2}\,D\right)\psi\,,
\end{gather}
is invariant under infinitesimal axial chiral transformations of the form~\cite{Luscher:1998pqa}
\begin{gather}
\label{eq:chiral}
\delta\bar\psi(x) = i\bar\psi(x)\gamma_5\,,~~~~~~~~~~
\delta\psi(x) = -i\gamma_5\tilde\psi(x)\,.
\end{gather}
Furthermore, all composite operators transform under~\req{eq:chiral}
as their continuum counterparts do under standard chiral transformations, provided
all quark fields $\psi$ are replaced by the rotated field $\tilde\psi$.
All the properties discussed above that make use of exact chiral symmetry thus
carry over to the regularised theory. One important technical issue is that local
conserved currents such as $\bar\psi\gamma_\mu\tilde\psi$ and $\bar\psi\gamma_\mu\gamma_5\tilde\psi$
still require a non-trivial finite normalisation with a constant $\ZV=\ZA$,
such that the correct chiral Ward-Takahashi identities hold.

\begin{figure}[t!]
\begin{center}
\includegraphics[width=100mm]{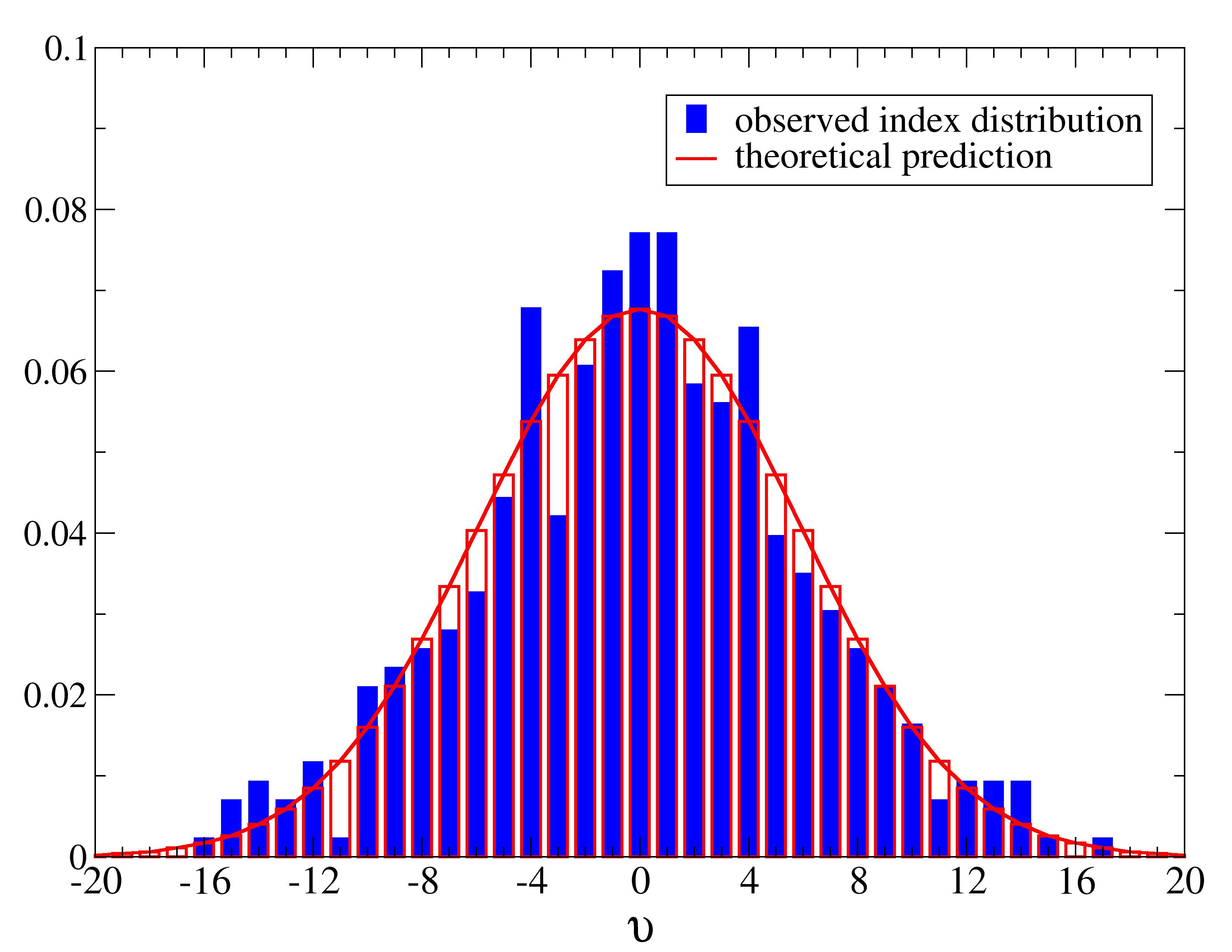}
\end{center}
\vspace{-5mm}
\caption{(Normalised) distribution of the index $\nu$ of the Neuberger-Dirac operator in the
gauge ensemble used for our $\epsilon$-regime computations (blue histogram), compared to the gaussian
shape expected in infinite volume (in red). The width of the gaussian has been computed with the value
of the topological susceptibility $r_0^4\chi = 0.00715(22)$ from~\cite{DelDebbio:2004ns}.
Note that the comparison does thus not involve any free parameter.}
\label{fig:index}
\end{figure}

Finally, one last crucial property of the Neuberger-Dirac operator is that
its index $\nu$ in a given gauge field provides a solid definition of the topological charge
associated to the latter~\cite{Hasenfratz:1998ri,Luscher:1998pqa}.
Thus, by computing zero modes of $D_{\rm N}$
one can split gauge ensembles into topological sectors in a well-defined
way. In~\refig{fig:index} we show the distribution
of topological charges for the ensemble used in our $\epsilon$-regime computations,
where correlation functions will be computed at fixed $\nu$.

Our simulations are carried out in the quenched approximation,
on a single lattice of size $32\times 16^3$ at $\beta=5.8485$.
This corresponds to a lattice spacing given, in
terms of the Sommer parameter $r_0 \approx 0.5~\fm$, by $a/r_0 \simeq 0.247$~\cite{Necco:2001xg}.
We always consider degenerate values of the light quark masses, $m_u=m_d=m_s\equiv m_l$.
Our simulation points are given in the first two columns of~\ret{tab:results_ratios}.
For the light masses we consider one $\epsilon$-regime point
($am_l=0.002$) and two $p$-regime points ($am_l=0.020,0,030$); the pseudoscalar octet meson
masses for the latter, measured from the two-point function of the non-singlet left-handed current,
are quoted in the third column of~\ret{tab:results_ratios}. For each light point then
we consider a value $m_c=m_l$, corresponding to the GIM limit, and two heavier charm
masses $am_c=0.040$ and $am_c=0.200$; for $am_l=0.020$ we also consider an even
heavier mass $am_c=0.400$.\footnote{Note that our simulation points in the GIM limit
coincide with some of the ones considered in~\cite{Giusti:2006mh}, which allows for a crosscheck of our
(independent) simulations.}
The value $am_c=0.040$ is still expected to be
within the reach of ChiPT, thus lying in the validity range of the study in~\cite{Hernandez:2006kz}.
Taking $r_0=0.5~\fm$ and the value of $a/r_0$ quoted before, our $p$-regime light pseudoscalar
meson masses correspond to $m_{\rm PS} \approx 317~\MeV$ and $m_{\rm PS} \approx 371~\MeV$.
Using also the value $\widehat Z_{\rm S}\simeq 1.28$ for the RGI scalar renormalisation constant
from~\cite{Wennekers:2005wa}, our three RGI charm masses for simulations at $m_c \neq m_l$
correspond, respectively, to $M_c\approx 50$, $249$, and $498~\MeV$. Note that, while the scaling
properties of computations with overlap fermions are generally expected to be good, at the heaviest
charm mass $am_c=0.400$ cutoff effects can be expected to be sizeable.

For each of the three values of $m_l$ we have an independent ensemble of around 400 independent
gauge configurations.
Only about half the statistics is used for the computation at $am_c=0.4$, as well as in the computation of three-point functions involving $Q_2^\pm$.

\subsection{Variance reduction techniques}

\begin{figure}[t!]

	\hspace*{-0mm}
    \begin{minipage}{.5\linewidth}
      \begin{center}
\begin{tikzpicture}[line width=0.6 pt, scale=0.6]
   \begin{scope}
   \draw[fermionbar] (0, 0) arc (0:180:1.4 and 1); 
   \draw[fermionbar]  (0, 0) arc (180:360:1.4 and 1); 
   \draw[fermion]     (0, 0) arc (0:180:1.4 and   -1); 
   \draw[fermion]     (0, 0) arc (180:360:1.4 and -1); 
    \draw[fill=black] (-2.8,0) circle (.15cm);
    \draw[fill=black] (2.8,0)  circle (.15cm);   
   \node at (-4.3,0.0) {$[J_{0}^{}(x)]_{du}^{}$}; 
   \node at ( 0.0,1.2) {$Q_{1}^{\pm}(z)$};
   \node at ( 4.5,0.0) {$[J_{0}^{}(y)]_{us}^{}$};  
   \draw[fill=black] (0,0) circle (.3cm);
    \draw[fill=white] (0,0) circle (.29cm);
    \begin{scope}
      \clip (0,0) circle (.3cm);
      \foreach \x in {-.9,-.8,...,.3}
    \draw[line width=1 pt] (\x,-.5) -- (\x+.6,.5);
    \end{scope}
   \end{scope}
\end{tikzpicture}
      \end{center}
    \end{minipage}
    \begin{minipage}{.5\linewidth}
      \vspace{-3.3mm}
      \begin{center}
 \begin{tikzpicture}[line width=0.6 pt, scale=0.55]
    \begin{scope}
  \draw[
        decoration={markings, mark=at position 0.75 with {\arrow{<}}},
        postaction={decorate}
        ] (0,-1) circle (1.0);
   \draw[fermionbar]  (0, 0) arc (90:180:2.4 and 1.6); 
   \draw[fermionbar]  (-2.4, -1.6) arc (180:360:2.4 and 1.6); 
   \draw[fermion]     (0, 0) arc (270:360:2.4 and -1.6); 
    \draw[fill=black] (-2.4,-1.6) circle (.15cm);
    \draw[fill=black] (2.4,-1.6)  circle (.15cm);   
   \draw[fill=black] (0,0) circle (.3cm);
    \draw[fill=white] (0,0) circle (.29cm); 
   \node at (-4.3,-1.6) {$[J_{0}^{}(x)]_{du}^{}$}; 
   \node at ( 0.0,0.8) {$Q_{1}^{\pm}(z)$};
   \node at ( 4.5,-1.6) {$[J_{0}^{}(y)]_{us}^{}$};
   \node at ( 0,-1.) {$\text{u/c}$};
   \begin{scope}
      \clip (0,0) circle (.3cm);
      \foreach \x in {-.9,-.8,...,.3}
    \draw[line width=1 pt] (\x,-.7) -- (\x+.6,.7);
    \end{scope}
   \end{scope}\label{fig:eye}
\end{tikzpicture}
      \end{center}
    \end{minipage}

\caption{Eight and eye diagrams appearing in the computation of three-point functions of $Q_1^\pm$.}
\label{fig:eight_eye}
\end{figure}
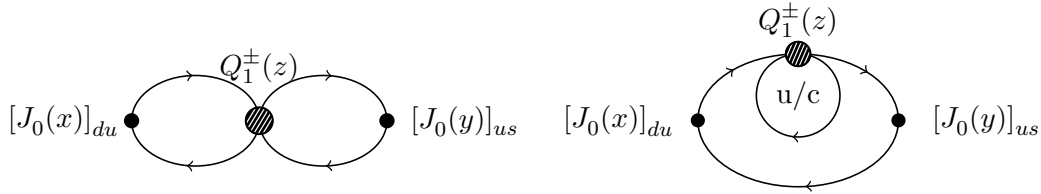

Our main aim is to compute the two- and three-point functions involved in the
matching of QCD to ChiPT, as discussed in \res{sec:strat}.
After integrating over fermion variables in the path integral,
fermionic correlation functions can be written as usual in terms
of gauge expectation values of traces of products of quark propagators
and spin matrices; explicit expressions are provided in~\reapp{app:traces}.
The reason to consider left-handed currents as interpolating operators
becomes apparent in that the traces only contain left-handed propagators
$P_-S(x,y)P_+$, that can always be computed in the chirality sector that does not
contain zero modes, thus avoiding their contribution in correlators~\cite{Giusti:2002sm}.
The three-point functions involving $Q_1^\pm$ require the computation
of the quark-propagator diagrams depicted in~\refig{fig:eight_eye},
to which we will refer as ``eight'' and ``eye'' diagrams, respectively.
Each of them appears in a colour-spin connected and a colour-spin disconnected
version.

The computation of these correlation functions poses severe problems
in terms of noise-to-signal ratio. When the light quark mass is sufficiently low
(and especially so in the $\epsilon$-regime), Dirac modes with very small eigenvalues
have large contributions to correlation functions. Their wavefunctions have been
shown to develop localised structures~\cite{Giusti:2003gf},
which makes good sampling of the whole lattice volume mandatory in order to
avoid large statistical fluctuations. It is thus important to integrate over
space at all operator insertion points (or at least at as many insertions as possible),
which obviously cannot be achieved with propagators computed with point sources.
The use of all-to-all propagators for variance reduction thus becomes mandatory.

One first step in this direction was the development of low-mode averaging (LMA)
in~\cite{DeGrand:2004qw,Giusti:2004yp}. In this particular brand of LMA the Dirac propagator $S$ is split into
the contribution $S_l$ from the lowest-lying $N_{\rm low}$ modes, which are treated exactly,
and its orthogonal complement $S_h$, which is computed with a point source.
This in turn implies a split of correlation functions into $2^{\rm L}$ different contributions,
where ${\rm L}$ is the number of propagators involved.
Contributions to correlation functions where two low propagators meet at
an operator insertion point can be integrated over space, since $S_l$ is effectively an all-to-all propagator.
On top of that, extra inversions performed using low modes as sources allow to
integrate also at insertions where one $S_l$ and one $S_h$ meet. This was
exploited in~\cite{Giusti:2004yp} to determine chiral LECs in the $\epsilon$-regime,
and in~\cite{Giusti:2005pd,Giusti:2006mh} to determine the weak LECs $g_1^\pm$ in the GIM $m_c=m_u$ limit,
implying that the noise-to-signal problem for eight diagrams is tamed via LMA.

The same techniques are however insufficient when applied to eye contractions;
in particular, the LMA technique does not allow to integrate over space at the insertion of
the four-fermion operator when $S_h$ circulates in the closed loop. One thus needs
to combine LMA with other variance reduction techniques, such as
stochastic volume sources (SVS)~\cite{Bernardson:1993he,Dong:1993pk},
and the novel probing algorithm proposed in~\cite{JTangYS}; the latter
can be used specifically for the
precise computation of closed propagators. A thorough study of these techniques
applied to our problem has been conducted in a companion paper~\cite{techpaper},
where the very large impact on variance reduction, at an affordable computational
cost, has been demonstrated.
In the present work, we have employed the optimised combination of LMA with SVS
developed in~\cite{techpaper}, to which we refer for full details,
with the specific aim of obtaining a well-behaved signal for the eye diagram.
The specific setup employed here treats the $20$
lowest modes of the Dirac operator exactly, and estimates $S^h$ with SVS using time and spin-colour
dilution and two stochastic hits.

In the case of the three-point function involving $Q_2^\pm$, a contribution from the
spin-diagonal part of the operator is unavoidable, since the presence of (pseudo)scalar
densities implies that not all propagators are left-handed. LMA has not been implemented
for these diagrams, and the only variance reduction techniques we employ for them is the
used of extended propagators, which allows to integrate over space at two of the three
operator insertions.
On the other hand, for this correlation function the prediction in~\req{eq:q2_chipt}
is expected to be accurate up to small NLO ChiPT corrections for light quark masses
in the $p$-regime; we can thus use the latter, together with the numerical results,
to provide a solid estimation.

\subsection{Results for ratios of correlation functions}

\begin{figure}[t!]

      \begin{center}
      \includegraphics[width=100mm]{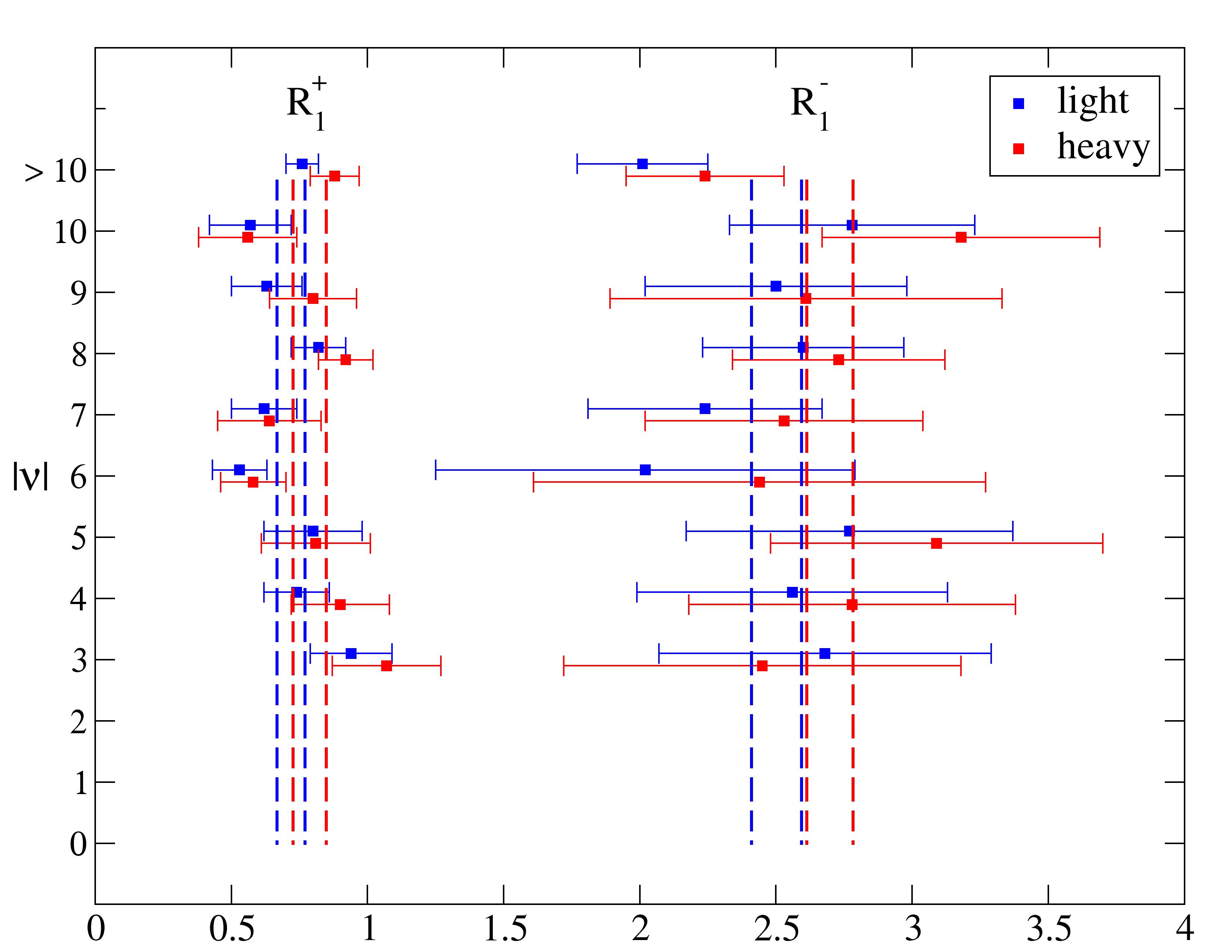}
      \end{center}

\caption{Values of the ratios $R_1^\pm$ in the $\epsilon$-regime as a function of $|\nu|$. ``Light'' and ``heavy'' refer to results for $am_c=0.040$ and $am_c=0.200$, respectively.}
\label{fig:eps}
\end{figure}

Sufficiently far away from operator insertions, the ratios involved in the matching to ChiPT
can be fitted to a plateau ansatz
so that correlation functions are dominated by the contribution from the lightest state.
Details about the fits are provided in~\reapp{app:fits};
our final results are quoted in~\ret{tab:results_ratios}.
Ratios in the $\epsilon$-regime are first computed in a fixed topological
sector $|\nu|$, and then a weighted average of the results for various values
of $|\nu|$ is taken. This procedure is based on the ChiPT prediction that the
ratios are insensitive to the value of $|\nu|$ up to NNLO corrections.
The results in~\ret{tab:results_ratios} include the topological sectors $3 \leq |\nu| \leq 7$.
This choice takes into account that no signal for eye diagrams is found for $|\nu|<3$,
and considering $|\nu|>7$ can be expected to introduce large finite volume effects.\footnote{The improvement
of the signal-to-noise ratio for this observable as $|\nu|$ increases had already been observed in~\cite{Giusti:2006mh,Hernandez:2008ft},
and is likely related to the fact that localised Dirac modes with small eigenvalues become less frequent
as the topological charge increases.}
\refig{fig:eps} illustrates the $|\nu|$
dependence of our results. The number of gauge configurations in the averages
for each value of $|\nu|$ is $\{42,57,36,29,25\}$, respectively.

\begin{table}[t!]
\begin{center}
\begin{tabular}{ccccc@{\hspace{2mm}}c@{\hspace{2mm}}c}
\Hline
$(am_l\,,\,am_c)$ & $am_{\rm PS}$ & $R_1^+$ & $R_1^-$ & $R_u^+$ & $R_2^\pm$ & $R_2^{\pm;{\rm ChiPT}}$ \\
\hline
0.002\,,\,0.002 &    ---     & 0.629(77) & 2.09(25) & 0.503(62) & 0 & 0 \\
0.002\,,\,0.040 &    ---     & 0.686(78) & 2.46(16) & 0.503(62) & n/a & -0.51(19) \\
0.002\,,\,0.200 &    ---     & 0.73(12)  & 2.68(13) & 0.503(62) & n/a & -13(4) \\ \hline
0.020\,,\,0.020 & 0.1986(20) & 0.692(25) & 1.972(63) & 0.554(20) & 0 & 0 \\
0.020\,,\,0.040 & 0.1986(20) & 0.717(25) & 2.028(64) & 0.554(20) & -0.36(12) & -0.38(7) \\
0.020\,,\,0.200 & 0.1986(20) & 0.766(32) & 2.220(82) & 0.554(20) & -12(4) & -13(3) \\
0.020\,,\,0.400 & 0.1986(20) & 0.767(51) & 2.42(12)  & 0.554(20) & -48(16) & -51(9) \\ \hline
0.030\,,\,0.030 & 0.2322(19) & 0.731(22) & 1.829(64) & 0.585(18) & 0 & 0 \\
0.030\,,\,0.040 & 0.2322(19) & 0.746(22) & 1.852(64) & 0.585(18) & n/a & -0.22(4) \\
0.030\,,\,0.200 & 0.2322(19) & 0.835(31) & 1.953(82) & 0.585(18) & n/a & -13(3) \\
\Hline
\end{tabular}
\end{center}
\caption{Bare quark masses, light pseudoscalar meson masses, and results for the ratios of QCD correlation functions involved in the matching to ChiPT.}
\label{tab:results_ratios}
\end{table}

In the case of the ratio $R_2^\pm$,
numerical results are provided in~\ret{tab:results_ratios} for $am_l=0.020$ only. We also provide the
LO ChiPT prediction for all kinematical points in the $p$-regime, using~\req{eq:chipt_r2}
with the bare values of quark masses.
The central value is set using $Fr_0 = 0.275(6)$ from~\cite{Giusti:2008fz},
and a systematic uncertainty that mimics the impact of NLO corrections,
obtained by varying $Fr_0$ in the range $0.250 \lesssim Fr_0 \lesssim 0.300$, is assigned.
This is a fairly conservative error estimate, as shown by the discussion in~\reapp{app:chipt}.
The current normalisation factor $\ZA^2$ needed to make connection with the ChiPT
prediction (cf.~\res{sec:subt}) is $\ZA = 1.706(5)$, taken from~\cite{Giusti:2008fz}.
Finally, by assuming that~\req{eq:chipt_r2} remains valid in the $\epsilon$-regime,
we also provide estimates of $R_2^\pm$ for the point $am_l=0.002$.
This assumption can be argued to hold on the
basis of the smooth $m_l\to 0$ limit of the relevant ChiPT formula for $\cR_8$, that
provides the $\epsilon$-regime value. In order to allow for
possible larger NLO (finite volume) corrections in this case, we have doubled
the size of the error estimate.

For the simulation points where a direct comparison is possible, the ChiPT prediction
is remarkably consistent with lattice data, within the relatively large errors displayed
by both quantities. Decreasing these errors would require a dedicated variance reduction
study, similar to the one conducted for correlators involving four-fermion operators.
Since, on the other hand, the contribution of $R_2^\pm$ to physical amplitudes is
suppressed by the small subtraction coefficients $c^\pm$, as discussed above, the level of precision
displayed by our results for $R_2^\pm$ in~\ret{tab:results_ratios} is good enough for
the purpose of the present work. We will henceforth take as input the values in the
last column of~\ret{tab:results_ratios} in the construction of the subtracted amplitudes.

In~\ret{tab:results_subtraction} we provide results for the ratios of the correlation
functions $D_1^\pm/D_2^\pm$ introduced in~\req{eq:subt_2p}, which are expected to exhibit
plateaux that can be fitted for the subtraction coefficients $c^\pm$. As explained
in section~\ref{sec:subt}, the dominant contribution in the $p$-regime comes from
scalar-to-vacuum amplitudes, which makes this quantity very noisy --- indeed no signal
is found from our data. The same applies to the ratios computed in the $\epsilon$-regime,
where the correlation functions do receive contributions from the pseudoscalar channel
but the intrinsic statistical fluctuations are also larger. We are thus unable to provide
a solid non-perturbative estimate of subtraction coefficients.
On the other hand, the error intervals
we find are compatible with the expectation $c^\pm \sim \cO(\alpha_{\rm s}/(4\pi))$.

In order to treat the contribution from the subtraction safely, we thus proceed as
follows. Subtraction coefficients are treated as suggested by the one-loop analysis
of~\res{sec:subt} --- i.e. set to zero, with a systematic uncertainty
given by $\alpha^{\MSbar}_{\rm s}(1/a)/(4\pi)\sim 0.028$.
When used together with the estimate of the subtraction term
coming from ChiPT, this leads to a systematic uncertainty on renormalised $K\to\pi$ amplitudes,
that should safely cover the effect of subtractions.
As the charm mass increases, the total error becomes increasingly dominated by this uncertainty.
However, the relative error on the final result is still around
or below $20\%$ for $am_c \leq 0.2$, and becomes very large only for $am_c=0.4$.
Using the values of the renormalisation
factors from~\cite{Dimopoulos:2006ma} quoted in~\ret{tab:renorm}, this leads to the renormalised ratios
in~\ret{tab:results_renorm}, that can then be used for the matching to
ChiPT.

\begin{table}[t!]
\begin{center}
\begin{tabular}{ccr@{\hspace{1pt}}lr@{\hspace{1pt}}l}
\Hline
$(am_l\,,\,am_c)$ & $am_{\rm PS}$ & & ~~~~$c^+$ & & ~~~~$c^-$ \\
\hline
0.002\,,\,0.040 &    ---     & &0.05(4) & -&0.14(48) \\
0.002\,,\,0.200 &    ---     & &0.00(3) & -&0.01(3) \\ \hline
0.020\,,\,0.040 & 0.1986(20) & -&0.01(12) & -&0.08(10) \\
0.020\,,\,0.200 & 0.1986(20) &  &0.00(1) & &0.00(9) \\ \hline
0.030\,,\,0.040 & 0.2322(19) & &0.04(10) & &0.14(48) \\
0.030\,,\,0.200 & 0.2322(19) & &0.01(21) & &0.02(8) \\
\Hline
\end{tabular}
\end{center}
\caption{Numerical results for the subtraction coefficients $c^\pm$, obtained from the ratios of correlation functions in~\req{eq:subt_2p}.}
\label{tab:results_subtraction}
\end{table}

\begin{table}[t!]
\begin{center}
\begin{tabular}{c@{\hspace{10mm}}cll}
\Hline
$\sigma$ & $k_1^{\sigma;{\rm RGI}}$ & $\runfac_1^\sigma(\mu/\Lambda)\frac{Z_{11}^\sigma(\mu)}{\ZA^2}$ & ~~$\cZ_1^\sigma$ \\
\hline
 $+$ & 0.7080 & ~~~1.15(12)  & 0.81(8) \\
 $-$ & 1.9775 & ~~~0.561(61) & 1.11(12) \\
\Hline
\end{tabular}
\end{center}
\caption{Values of Wilson coefficients and renormalisation factors for quenched QCD at $\beta=5.8485$ (from~\cite{Dimopoulos:2006ma}).}
\label{tab:renorm}
\end{table}

\section{Matching to Chiral Perturbation Theory}
\label{sec:match}

In order to determine the values of $g_8(m_c)$ and $g_{27}(m_c)$, the renormalised QCD
quantities $R_{27},R_8$ in~\ret{tab:results_renorm} and the
ChiPT ratios $\cR_{27},\cR_8$ in~\reapp{app:chipt}
have to be introduced into~Eqs.~(\ref{eq:match27},\ref{eq:match8}),
for each of the values of $m_c$ available, apart from $am_c=0.4$
--- for which we have results only at one value of the light mass,
and errors are large.
As already noted, our results in the GIM limit are well-consistent with those in~\cite{Giusti:2006mh}
for the same simulations points --- differences are always below the $2\sigma$
level.\footnote{Ideally, one would like to perform independent fits in the $\epsilon$- and $p$-regime;
consistent results would then indicate that higher-orders ChiPT corrections are well under control,
and a simultaneous fit of both regimes can be used to obtain definitive results for the LO LECs.
This was indeed the strategy successfully pursued in~\cite{Giusti:2006mh}. In this work, however, having only two $p$-regime
masses does not allow for meaningful fits involving $p$-regime points only, and therefore we will
only quote results coming from combined fits. The study in~\cite{Giusti:2006mh} supports the underlying
assumption that higher-order effects are adequately covered by our errors.}

\begin{table}[t!]
\begin{center}
\begin{tabular}{cccc}
\Hline
$(am_l\,,\,am_c)$ & $am_{\rm PS}$ & $R_{27}$ & $R_8$ \\
\hline
0.002\,,\,0.002 &    ---     & 0.407(69) & 2.42(38) \\
0.002\,,\,0.040 &    ---     & 0.407(69) & 2.88(35) \\
0.002\,,\,0.200 &    ---     & 0.407(69) & 3.16(62) \\ \hline
0.020\,,\,0.020 & 0.1986(20) & 0.449(48) & 2.30(25) \\
0.020\,,\,0.040 & 0.1986(20) & 0.449(48) & 2.38(26) \\
0.020\,,\,0.200 & 0.1986(20) & 0.449(48) & 2.64(58) \\
0.020\,,\,0.400 & 0.1986(20) & 0.449(48) & 2.9(2.0) \\ \hline
0.030\,,\,0.030 & 0.2322(19) & 0.474(50) & 2.15(23) \\
0.030\,,\,0.040 & 0.2322(19) & 0.474(50) & 2.19(23) \\
0.030\,,\,0.200 & 0.2322(19) & 0.474(50) & 2.36(56) \\
\Hline
\end{tabular}
\end{center}
\caption{Renormalised QCD ratios that enter the matching to ChiPT.}
\label{tab:results_renorm}
\end{table}

A straightforward procedure follows by rewriting~Eqs.~(\ref{eq:match27},\ref{eq:match8})
as
\begin{gather}
\begin{split}
R_{27}(m_l,m_c) &= g_{27}(m_c)\left\{1-\Delta_{27}[m_l,\Lambda_{27}(m_c)]\right\}\,,\\
R_{8}(m_l,m_c) &= g_{8}(m_c)\left\{1-\Delta_{8}[m_l,\Lambda_{8}(m_c)]\right\}\,,
\end{split}
\end{gather}
where for greater clarity we have made quark mass dependences explicit.
Here $\Delta_k(m_l)$ is either the NLO (finite-volume) correction
in the $\epsilon$-regime (for which effectively $m_l=0$),
\begin{gather}
\Delta_{27}^\epsilon = 0.182(8)\,,~~~~~~~~~~~~~
\Delta_8^\epsilon = -0.273(12)\,.
\end{gather}
or the $p$-regime correction involving
chiral logs plus finite-volume terms.\footnote{The latter
are anyway expected to be small in our case --- in our simulations the parameter
that controls finite-volume corrections is $\sim \exp(-m_{\rm PS}L) \lesssim 0.04$.}
The scales $\Lambda_k$ parametrise contributions from NLO terms in the $p$-regime
chiral effective Hamiltonian.
By setting $Fr_0=0.275(6)$, one can then fit our three $m_l$ data points,
separately in the 27-plet and octet channels and for each value of $m_c$,
to determine the two parameters $g_k(m_c)$ and $\Lambda_k(m_c)$.
Note that all the data points come from different gauge ensembles,
which makes their correlation negligible.

\begin{figure}[t!]
\begin{center}
\includegraphics[width=100mm]{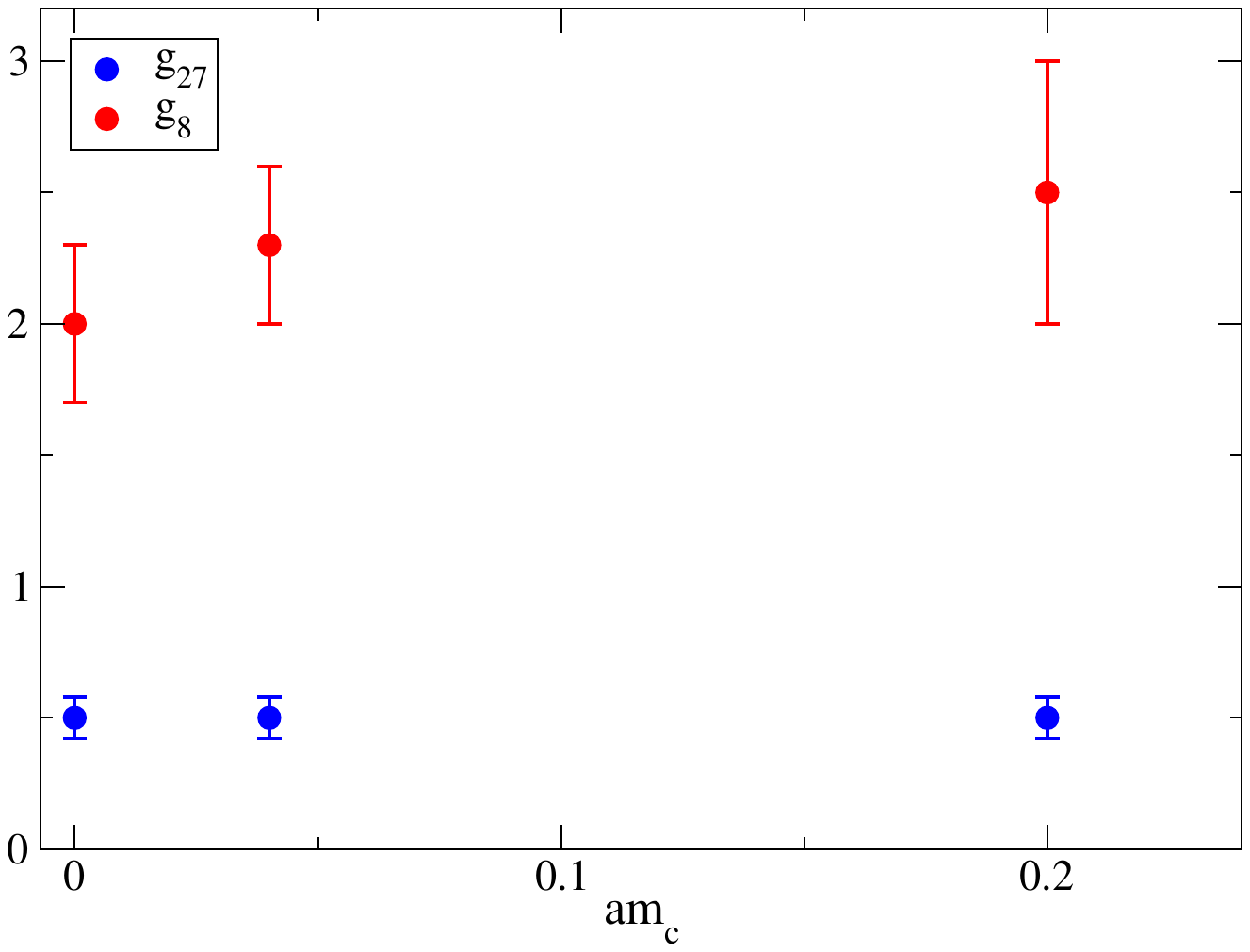}
\end{center}
\vspace{-10mm}
\caption{LO LECs $g_{27}$ and $g_8$ as a function of $am_c$. Recall the ``experimental'' values
$g_{27} \sim 0.50$ and $g_8 \sim 10.5$ (cf.~\res{sec:strat}).}
\label{fig:LECs}
\end{figure}

As discussed in~\res{sec:strat}, the matching to ${\rm SU}(3)$ ChiPT of quenched results
is problematic in the octet case. In particular, singlet contributions to the formulae
in~\reapp{app:chipt} should be taken into account. Since, on the other hand, the errors
on $R_8$ are large, and we only have results at two $p$-regime quark masses, the sensitivity
to these NLO effects is very poor. We have fit our numbers to the
$N_{\rm f}=2$, $N_{\rm f}=3$, and $N_{\rm f}=4$ formulae, and find that the value
of $g_8$ is completely insensitive to $N_{\rm f}$; only $\Lambda_8$ changes, as shown
in~\ret{tab:fits_lecs}.
The result we thus quote for the LO LECs is
\begin{gather}
\label{eq:fit1}
\begin{array}{l@{\hspace{10mm}}r@{\hspace{1mm}}c@{\hspace{1mm}}l@{\hspace{5mm}}r@{\hspace{1mm}}c@{\hspace{1mm}}l}
am_c = 0.00: & g_{27} & = & 0.50(8)\,, & g_{8}  & = & 1.9(3)\,; \\[2.0ex]
am_c = 0.04: & g_{27} & = & 0.50(8)\,, & g_{8}  & = & 2.3(3)\,; \\[2.0ex]
am_c = 0.20: & g_{27} & = & 0.50(8)\,, & g_{8}  & = & 2.5(5)\,, \\
\end{array}
\end{gather}
where we have also included (labeling it as $m_c=0$) the result of a reanalysis of the GIM limit based
on our simulations. The latter is again consistent within $\sim 1\sigma$ with the conclusions
in~\cite{Giusti:2006mh}. Recall that, since we are working in the quenched approximation,
the LEC $g_{27}$ is strictly independent of $m_c$.
These fit results are illustrated in~\refig{fig:LECs}.

\begin{table}[t!]
\begin{center}
\begin{tabular}{cccc}
\Hline
$\NF$ & $am_c$ & $g_8$ & $\Lambda_8$ \\
\hline
2 & 0.00 & 1.92(28) & 0.28(9) \\
3 & 0.00 & 1.94(28) & 0.32(16) \\
4 & 0.00 & 1.94(29) & 0.37(26) \\ \hline
2 & 0.04 & 2.26(26) & 0.22(5) \\
3 & 0.04 & 2.28(26) & 0.22(8) \\
4 & 0.04 & 2.59(11) & 0.26(11) \\ \hline
2 & 0.20 & 2.49(47) & 0.22(9) \\
3 & 0.20 & 2.50(48) & 0.21(14) \\
4 & 0.20 & 2.50(48) & 0.21(20) \\ \hline
\Hline
\end{tabular}
\end{center}
\caption{Results of fits to ChiPT formulae for $g_8$ and $\Lambda_8$. (See text for an explanation
of the $\NF$ dependence of the fit function; the $\epsilon$-regime point is labeled $am_c=0.00$;
the (correlated) $\chi^2/{\rm d.o.f.}$ of the fits is always $\lesssim 10^{-2}$.) }
\label{tab:fits_lecs}
\end{table}

Alternatively, as discussed in~\cite{Giusti:2006mh}, fits can be performed
to the product $R_{27}R_8$, which is less sensitive to chiral corrections,
and take the value of $g_{27}(0)$ as input from the more solid determination
in that work (which has better $\epsilon$-regime statistics and additional $p$-regime masses).
The fit ansatz for the product of ratios is
\begin{gather}
R_{27}\,R_8 = g_{27}g_8[1-\tilde\Delta]\,,
\end{gather}
where $\tilde\Delta=\Delta_{27}+\Delta_8$ --- explicitly
\begin{gather}
\begin{split}
\tilde\Delta^\epsilon &= -0.091(4)\,,\\
\tilde\Delta^p &= -\frac{M^2}{(4\pi F)^2}\left[
\left(8+\frac{10}{N_{\rm f}}\right)\log\left(\frac{M^2}{\tilde\Lambda^2}\right) -(\cV_{27}+\cV_8)
\right]\,,
\end{split}
\end{gather}
where $\tilde\Lambda$ is a single scale that combines the effect of NLO
terms in the 27-plet and octet channel (cf.~\reapp{app:chipt} for unexplained notation).
We follow the same procedure to check the dependence on $N_{\rm f}$ as before,
finding similar results.
The outcome of this latter fit strategy is
\begin{gather}
\label{eq:fit2}
\begin{array}{l@{\hspace{10mm}}r@{\hspace{1mm}}c@{\hspace{1mm}}l}
am_c = 0.00: & g_{27}g_8 & = & 0.94(2)\,; \\[1.0ex]
am_c = 0.04: & g_{27}g_8 & = & 1.10(2)\,; \\[1.0ex]
am_c = 0.20: & g_{27}g_8 & = & 1.20(4)\,,
\end{array}
\end{gather}
which exhibits good consistency with the results in \req{eq:fit1},
and checks that they are robust.



\section{Conclusions}
\label{sec:concl}

In this paper we have explored the behaviour of the $K\to\pi\pi$ decay amplitudes
involved in the $\Delta I=1/2$ rule as a function of the charm quark mass, following
the strategy laid out in~\cite{Giusti:2004an}. The aim is to understand the role of the charm quark
in the $\Delta I=1/2$ enhancement. Our work extends the results for the GIM limit $m_c=m_u$
in~\cite{Giusti:2006mh,Hernandez:2008ft}. The numerical techniques developed in~\cite{techpaper}
have been instrumental in the lattice QCD computation of amplitudes involving eye diagrams.

Our main finding is that unsubtracted matrix elements of the four-fermion
operators $Q_1^\pm$, computed in quenched QCD, have a mild dependence on the charm-up quark mass 
difference across the regime where the charm quark becomes heavy. Indeed, while our simulations
do not reach the physical value of the charm mass, they cover values of $m_c$ about 100 times 
larger than the physical value of $(m_u+m_d)/2$.
At that point, the dominant contribution to the enhancement from $Q_1^-$ increases
by no more than $\sim 30\%$ with respect to the value found with light and mass-degenerate
up and charm quarks.

We have also discussed how the subtraction term needed to obtain
the physical amplitudes for $m_c \neq m_u$ is proportional to $m_c^2$ for a heavy charm.
Combined with the above result, this would imply that the ratio of low-energy couplings $g_8/g_{27}$
is bound to become large as the charm mass increases, since the contribution from the
subtraction term will eventually dominate.
Alternatively, bare matrix elements of $Q_1^\pm$ may start showing
a larger $m_c$ dependence closer to the physical charm mass value, allowing for potential cancellations.
This however seems unnatural, since, as pointed above, our $m_c$ values are already well
above the light quark regime.
In that sense, our results point in the direction of supporting that a strong
enhancement is natural for large enough values of $m_c/m_u$.

On the other hand, our results are insufficient to determine the contribution
from the subtraction term precisely. While the value of the matrix elements of the operator
$Q_2^\pm$ involved in the subtraction are well-controlled (within sizeable uncertainties),
further work is needed for a reliable non-perturbative determination of the
subtraction coefficients $c^\pm$. In the interpretation that the $m_c$ dependence
at large $m_c$ is driven by the subtraction term, the value of $c^\pm$ is crucial
to fix the precise value of $|A_0|/|A_2|$ at the physical point. Assuming the suppression
in $c^\pm$ hinted at by perturbation theory, we have found that the enhancement already
observed in the GIM limit does not increase significantly within the range of values of $m_c$
covered by our simulations. The ultimate question whether Standard Model physics
alone can quantitatively explain the experimental value of $|A_0|/|A_2|$ is thus left open ---
answering it within our framework still requires a more detailed study of the subtraction terms,
as well as reaching out to values of the charm mass in the physical region.

\section*{Acknowledgments}

CP is indebted to Leonardo Giusti, Pilar Hern\'andez, Mikko Laine,
Jan Wennekers, and Hartmut Wittig for many illuminating discussions in the
context of the approach to the $\Delta I=1/2$ rule of which this work makes part.
We would especially like to thank Pilar Hern\'andez for reading the manuscript
and making valuable suggestions, and Margarita Garc\'{\i}a P\'erez and
Tassos Vladikas for several discussions on the topics covered here.
Our numerical computations have been carried out at the
Altamira and MareNostrum installations of the Spanish Supercomputation Network,
the Hydra cluster at IFT, and the Finisterrae installation at CESGA. Support
by the staff from these centers is gratefully acknowledged.
This work has
been supported by the Spanish MICINN under grant FPA2009-08785, the Spanish MINECO under grant
FPA2012-31686 and the ÓCentro de excelencia Severo Ochoa ProgramÓ SEV-2012-0249, the
Community of Madrid under grant HEPHACOS S2009/ESP-1473, and especially by the European Union
under the Marie Curie-ITN Program STRONGnet, grant PITN-GA-2009-238353.

\newpage
\begin{appendix}
\section{Chiral Perturbation Theory formulae}
\label{app:chipt}

In this appendix we collect the essential next-to-leading order quenched ChiPT formulae
from~\cite{Giusti:2004an,Hernandez:2006kz} relevant for the determination of the
LECs in the ${\rm SU}(4)$ and ${\rm SU}(3)$ chiral effective weak Hamiltonians.
We also discuss NLO ChiPT corrections to the ratio $(m_K^2-m_\pi^2)/(m_s-m_d)$ that
determines matrix elements of $Q_2^\pm$ in our kinematics.

\subsection{NLO corrections to chiral weak Hamiltonians}

Here we provide NLO results for the various ratios of correlation functions
in ChiPT discussed in the text, taken from~\cite{Hernandez:2006kz}.
Note that $\epsilon$-regime results are given for a specific topological sector
with topological charge $\nu$. In particular, the ratios $\cR_1^\pm,\cR_{27},\cR_8$
happen to be independent of $\nu$ up to NNLO corrections, while the expressions
involving the unphysical operator $\cO'_8$ do exhibit topology dependence,
but they are not included here since their explicit form is not needed in the matching.
In $p$-regime expressions, the contributions from unknown NLO LECs are
included in the scales appearing in chiral logarithms.
For ${\rm SU}(3)$ ChiPT in the octet channel we
quote the unquenched formulae; comments about the matching to quenched QCD results
are provided in~\res{sec:strat} and~\res{sec:match}.

All equations hold in a box with four-volume $V=L^3 \times T$
and aspect ratio $\rho = T/L$. The dependence on the light quark mass $m_l$ is given
either in terms of the leading-order Goldstone boson mass $M^2=2\Sigma m_l/F^2$ ($p$-regime),
or in terms of the dimensionless parameter $\mu=m_l\Sigma V$ ($\epsilon$-regime).

\paragraph{${\rm SU}(4)$ ChiPT, $\epsilon$-regime:}

\begin{gather}
\cR_{1;\nu}^\pm(x_0,y_0) = 1 \pm\,\frac{2}{F^2 T^2}\,(\beta_1\rho^{3/2}-k_{00}\rho^3)\,.
\end{gather}

\paragraph{${\rm SU}(3)$ ChiPT, $\epsilon$-regime:}

\begin{align}
\cR_{27;\nu}(x_0,y_0) &= 1 +\,\frac{2}{F^2 T^2}\,(\beta_1\rho^{3/2}-k_{00}\rho^3)\,,\\
\cR_{8;\nu}(x_0,y_0) &= 1 -\,\frac{3}{F^2 T^2}\,(\beta_1\rho^{3/2}-k_{00}\rho^3)\,.
\end{align}

\paragraph{${\rm SU}(4)$ ChiPT, $p$-regime:}

\begin{gather}
\cR_1^\pm(x_0,y_0) = 1 \pm \frac{M^2}{(4\pi F)^2}\left[3\log\left(\frac{M^2}{\Lambda_\pm^2}\right)
\mp \cV_1(x_0,y_0)\right]\,.
\end{gather}

\paragraph{${\rm SU}(3)$ ChiPT, $p$-regime:}

\begin{align}
\cR_{27}(x_0,y_0) &= 1 + \frac{M^2}{(4\pi F)^2}\left[3\log\left(\frac{M^2}{\Lambda_{27}^2}\right)
- \cV_{27}(x_0,y_0)\right]\,,\\
\cR_{8}(x_0,y_0) &= 1 + \frac{M^2}{(4\pi F)^2}\left[\left(\frac{1}{2}-\frac{10}{N_{\rm f}}\right)\log\left(\frac{M^2}{\Lambda_8^2}\right)
- \cV_8(x_0,y_0)\right]\,.
\end{align}

\paragraph{Finite volume effects:}

NLO corrections in the $\epsilon$-regime are pure finite-volume effects,
parametrised by the geometrical coefficients~\cite{Hasenfratz:1989pk,Hansen:1990un,Hansen:1990yg}
\begin{align}
\beta_1 &= \frac{1}{4\pi}\left[2-\hat\alpha_{-1}(\rho^{3/4},\rho^{-1/4})-\hat\alpha_{-1}(\rho^{-3/4},\rho^{1/4})\right]\,,\\
k_{00} &= \frac{1}{12}-\frac{1}{4}\sum_{\vn\neq\mathbf{0}}\frac{1}{\sinh^2(\pi\rho|\vn|)}\,,
\end{align}
where $\vn$ are integer vectors, and $\hat\alpha_p$ is given in terms of the elliptic theta function
$S(x)=\sum_{n=-\infty}^\infty \exp(-\pi x n^2) = \vartheta_3(0,\exp(-\pi x))$ by
\begin{gather}
\hat\alpha_p(l_0,l_i) = \int_0^1 \dif t\,t^{p-1}\left[S(l_0^2/t)S^3(l_i^2/t)-1\right]\,.
\end{gather}
A table with sample values of $\beta_1,k_{00}$ is provided in~Table~4 of~\cite{Giusti:2004an}.
In our lattice,
\begin{gather}
\beta_1 = 0.08360 \,,~~~~~~~~~~~~~~~~~~~~~
k_{00} = 0.08331.
\end{gather}
This implies, in particular, that the parameter that controls
$\epsilon$-regime NLO corrections is $(\beta_1\rho^{3/2}-k_{00}\rho^3)/(F^2T^2) \approx -0.13$,
taking $F \approx 90~\MeV$ and $T \approx 4~\fm$. That implies large corrections ranging between
$\sim 25\%$ and $\sim 40\%$ in the $\epsilon$-regime matching for LECs.

Finite-volume effects in $p$-regime ratios involving three-point functions are
given, in sufficiently large volumes, by
\begin{align}
\cV_1(x_0,y_0)=\cV_{27}(x_0,y_0) &= e^{-2M|x_0|}\phi_1(2M|x_0|) + e^{-2M|y_0|}\phi_1(2M|y_0|)\,,\\
\cV_8(x_0,y_0) &= e^{-2M|x_0|}\phi_2(2M|x_0|) + e^{-2M|y_0|}\phi_2(2M|y_0|)\,,
\end{align}
with
\begin{align}
\phi_1(x) &= \int_0^\infty \dif z\,e^{-xz}\frac{\sqrt{z(2+z)}}{1+z}\left[\frac{1}{2+z}+\frac{1}{1+z}-2\right]\,,\\
\phi_2(x) &= \int_0^\infty \dif z\,e^{-xz}\frac{\sqrt{z(2+z)}}{1+z}\left[
\frac{-4+\frac{5}{N_{\rm f}}}{2+z}+\frac{1-\frac{5}{N_{\rm f}}}{1+z}
-2+\frac{10}{N_{\rm f}}-\left(10-\frac{20}{N_{\rm f}}\right)z
\right]\,.
\end{align}
Note that the dependence of these quantities on $(x_0,y_0)$ is actually very mild; in fits
we will take their values at $x_0=-y_0=T/3$.

\subsection{NLO corrections to $(m_K^2-m_\pi^2)/(m_s-m_d)$}

The full NLO expression for the ratio $(m_K^2-m_\pi^2)/(m_s-m_d)$ is given by~\cite{Gasser:1984gg}
(we take $m_u=m_d$ throughout; general expressions can be obtained by replacing occurrences
of $m_d$ by $m_{ud}=\half(m_u+m_d)$)
\begin{gather}
\begin{split}
\frac{m_K^2-m_\pi^2}{m_s-m_d} =
\frac{\Sigma_0}{F_0^2}\bigg\{1 &+ \frac{\Sigma_0}{8\pi^2 F_0^4}\left[(m_s+m_d)\,\ell_1 + m_d\,\ell_2\right]\\
&+
\frac{8\Sigma_0}{F_0^4}\left[
(m_s+3m_d)(2L_8-L_5)+2(m_s+2m_d)(2L_6-L_4)
\right]
\bigg\}\,,
\end{split}
\end{gather}
where $F_0,\Sigma_0,L_i$ are the standard ${\rm SU}(3)$ LECs, and the logarithm terms $\ell_{1,2}$ read
\begin{align}
\ell_1 &= \frac{2}{9}\log\left[\frac{2\Sigma_0(2m_s+m_d)}{3\mu^2F_0^2}\right]\,,\\
\ell_2 &= \left(\frac{m_s}{m_s-m_d}\right)\log\left[\frac{2\Sigma_0(2m_s+m_d)}{3\mu^2F_0^2}\right]
-\left(\frac{m_d}{m_s-m_d}\right)\log\left[\frac{2\Sigma_0m_d}{\mu^2F_0^2}\right]\,.
\end{align}
Following standard practice, we assume $\mu=770~\MeV$ as the scale at which
the logarithms, quark masses, and NLO LECs $L_i$ are evaluated.
The term $\ell_2$ does not transparently have a well-behaved $m_s=m_d$ limit, but it is
easy to show that taking $m_s=m_d(1+\epsilon)$ one can write it as
\begin{gather}
\ell_2 = \frac{2}{3}+\log\left[\frac{2\Sigma_0 m_d}{\mu^2F_0^2}\right]\,+\,\cO(\epsilon)\,.
\end{gather}
The result for $m_s=m_d=m_u \equiv m_l$ simplifies to
\begin{gather}
\begin{split}
\frac{m_K^2-m_\pi^2}{m_s-m_d} \to \frac{\Sigma_0}{F_0^2}\bigg\{1
&+\frac{\Sigma_0m_l}{8\pi^2F_0^4}\left(\frac{2}{3}+\frac{13}{9}\log\left[\frac{2\Sigma_0m_l}{\mu^2F_0^2}\right]\right)\\
&+\frac{16\Sigma_0m_l}{F_0^4}\left[2(2L_8-L_5)+3(2L_6-L_4)\right]
\bigg\}\,.
\end{split}
\end{gather}
In the quenched case there will be additional contributions from the non-decoupled
singlet terms, which can be reabsorbed in a renormalised chiral condensate $\bar\Sigma_0$,
that will diverge in the chiral limit.

Current reference values for the relevant LECs, obtained from $N_f=2+1$ lattice
simulations, are~\cite{Bazavov:2009bb,Bazavov:2009fk,Bazavov:2010hj,Aoki:2013ldr}
\begin{gather}
\begin{split}
F_0 = 80(6)~\MeV\,,&~~~~~~~~~
\Sigma_0^{1/3} = 245(8)~\MeV\,,\\
(2L_6-L_4) = 0.03^{+40}_{-36} \times 10^{-3}\,,&~~~~~~~~~
(2L_8-L_5) = -0.15^{+46}_{-22} \times 10^{-3}\,.
\end{split}
\end{gather}
This implies $\Sigma_0/F_0^4 \simeq 0.36(11)~\MeV^{-1}$, and therefore a conservative
upper bound for the size of NLO corrections for values of $m_l \lesssim m_s^{\rm phys}/4$,
as is our case, can be taken to be $\sim 5\%$, which we increase to $\sim 10\%$
to account for deviations from this scenario in the quenched
case (which can be expected to be small, as shown by the values for LO quenched LECs derived from a
similar lattice setup to the one used in this work~\cite{Giusti:2008fz}).

\section{Wick contractions for QCD correlation functions}
\label{app:traces}

\subsection{Three-point functions of $Q_1^\pm$}

In the limit $m_u=m_d=m_s=m_l$, the QCD three-point functions involving $Q_1^\pm$ needed in our setup can be
computed in terms of a few independent fermionic traces.
Without loss of generality, we will write the expressions for a four-fermion operator
inserted at $z=0$.
Let $S_l(x,y)$ and $S_c(x,y)$ be
the propagators of a light quark and a charm quark, respectively, and let us define
\begin{align}
E^{\rm D}(x_0,y_0) &= \int_{\vx,\vy}\langle
{\rm Tr}\left[
S_l(x,0) \gamma_\mu P_- S_l(0,x) \gamma_0 P_-
\right]
{\rm Tr}\left[
S_l(y,0) \gamma_\mu P_- S_l(0,y) \gamma_0 P_-
\right]
\rangle_{\rm G}\,,\\
E^{\rm C}(x_0,y_0) &= \int_{\vx,\vy}\langle
{\rm Tr}\left[
S_l(x,0) \gamma_\mu P_- S_l(0,y) \gamma_0 P_- S_l(y,0) \gamma_\mu P_- S_l(0,x) \gamma_0 P_-
\right]
\rangle_{\rm G}\,,\\
P_l^{\rm D}(x_0,y_0) &= \int_{\vx,\vy}\langle
{\rm Tr}\left[
S_l(0,0) \gamma_\mu P_-
\right]
{\rm Tr}\left[
S_l(0,x) \gamma_0 P_- S_l(x,y) \gamma_0 P_- S_l(y,0) \gamma_\mu P_-
\right]
\rangle_{\rm G}\,,\\
P_l^{\rm C}(x_0,y_0) &= \int_{\vx,\vy}\langle
{\rm Tr}\left[
S_l(0,0) \gamma_\mu P_- S_l(0,x) \gamma_0 P_- S_l(x,y) \gamma_0 P_- S_l(y,0) \gamma_\mu P_-
\right]
\rangle_{\rm G}\,,\\
P_c^{\rm D}(x_0,y_0) &= \int_{\vx,\vy}\langle
{\rm Tr}\left[
S_c(0,0) \gamma_\mu P_-
\right]
{\rm Tr}\left[
S_l(0,x) \gamma_0 P_- S_l(x,y) \gamma_0 P_- S_l(y,0) \gamma_\mu P_-
\right]
\rangle_{\rm G}\,,\\
P_c^{\rm C}(x_0,y_0) &= \int_{\vx,\vy}\langle
{\rm Tr}\left[
S_c(0,0) \gamma_\mu P_- S_l(0,x) \gamma_0 P_- S_l(x,y) \gamma_0 P_- S_l(y,0) \gamma_\mu P_-
\right]
\rangle_{\rm G}\,,
\end{align}
where traces are taken over spin and colour indices, and $\langle\rangle_{\rm G}$ means that
the expectation value is taken in the pure Yang-Mills theory with the effective action
resulting from integration over quark fields in the path integral. Some straightforward
algebra then shows that all the three-point functions of the four-fermion operators considered
in the text with two left-handed currents can be written as
\begin{align}
C_1^+ &= [E^{\rm D} - E^{\rm C}] + [P_l^{\rm D} - P_c^{\rm D}] - [P_l^{\rm C} - P_c^{\rm C}] \,,\\
C_1^- &= [E^{\rm D} + E^{\rm C}] - [P_l^{\rm D} - P_c^{\rm D}] - [P_l^{\rm C} - P_c^{\rm C}] \,,\\
C_u^+ &= \frac{4}{5}\,[E^{\rm D} - E^{\rm C}] \,,\\
{\scriptstyle \frac{1}{5}}C_R^+ - C_c^+ &= {\scriptstyle \frac{1}{5}}[E^{\rm D} - E^{\rm C}] + [P_l^{\rm D} - P_c^{\rm D}] - [P_l^{\rm C} - P_c^{\rm C}] \,,\\
C_R^- - C_c^- &= [E^{\rm D} + E^{\rm C}] - [P_l^{\rm D} - P_c^{\rm D}] - [P_l^{\rm C} - P_c^{\rm C}] \,.
\end{align}

\subsection{Three-point functions of $Q_2^\pm$}

The three-point functions $C_2^\pm$ for the insertion of $Q_2^\pm$ at $z=0$ can be written as
\begin{gather}
C_2^\pm(x_0,y_0)= \frac{1}{2}(m_u^2-m_c^2)\left\{
(m_s+m_d)C_{\rm S}(x_0,y_0) - (m_s-m_d)C_{\rm P}(x_0,y_0)
\right\}\,,
\end{gather}
with
\begin{align}
C_{\rm S}(x_0,y_0) &= -\int_{\vx,\vy}\langle {\rm Tr}\left[S_l(0,x)\gamma_0 P_-S_l(x,y)\gamma_0 P_-S_l(y,0)\right]\rangle_{\rm G}\,,\\
C_{\rm P}(x_0,y_0) &= -\int_{\vx,\vy}\langle {\rm Tr}\left[S_l(0,x)\gamma_0 P_-S_l(x,y)\gamma_0 P_-S_l(y,0)\gamma_5\right]\rangle_{\rm G}\,.
\end{align}

\subsection{Two-point functions}

We consider two-point functions of a left-handed current (let us say at $y=0$)
with either another left-handed current, a scalar density, or a pseudoscalar density,
always in the light sector and in a non-singlet flavour channel.
The relevant Wick contractions are of the form
\begin{gather}
-\int_\vx \langle{\rm Tr}\left[
S_l(x,0)\gamma_0 P_- S_l(0,x)\Gamma
\right]\rangle_{\rm G}\,,
\end{gather}
where $\Gamma=\gamma_0 P_-,\mathbf{1},\gamma_5$ for each of the three possibilities mentioned above.

\section{One-loop study of subtraction coefficients}
\label{app:pert}

Our starting point is the subtraction condition in \req{eq:risubt}.
Substituting \req{eq:hren} into that expression one has
\begin{gather}
\label{eq:risubt_exp}
Z_{11}^\pm\left\{F_1^\pm\,+\,\half\,c^\pm\,(m_u^2-m_c^2)
\left[(m_s+m_d)F_{\rm S}-(m_s-m_d)F_{\rm P}\right]\right\}\,,
\end{gather}
where
\begin{gather}
\begin{split}
F_1^{\pm} &= {\rm tr}\langle s(p)\,Q_1^\pm\,\bar d(p) \rangle_{\rm amp}\,,\\
F_{\rm S} &= {\rm tr}\langle s(p)\,(\bar s d)\,\bar d(p) \rangle_{\rm amp}\,,\\
F_{\rm P} &= {\rm tr}\langle s(p)\,(\bar s \gamma_5 d)\,\bar d(p) \rangle_{\rm amp}\,.
\end{split}
\end{gather}
Each of these amputated correlation functions depends on the external momentum
$p$ and on the quark masses $m_i,~i=u,d,s,c$. After performing Wick contractions,
these correlators can be written as
\begin{gather}
\begin{split}
F_1^{\pm} &= \int\kern-2pt\frac{\dif^4q}{(2\pi)^4}\Big\{
{\rm tr}\langle
\tilde S_s(p)\gamma_\mu^{\rm L}\tilde S_{u-c}(q) \gamma_\mu^{\rm L}\tilde S_d(p)
\rangle_{\rm amp}\\
&~~~~~~~~~~~~~~~~\,\mp\,
{\rm tr}\langle
\tilde S_s(p)\gamma_\mu^{\rm L}\tilde S_d(p) {\rm tr}[\gamma_\mu^{\rm L} \tilde S_{u-c}(q)]
\rangle_{\rm amp}
\Big\}\,,\\
F_{\rm S} &= {\rm tr}\langle \tilde S_s(p)\tilde S_d(p) \rangle_{\rm amp}\,,\\
F_{\rm P} &= {\rm tr}\langle \tilde S_s(p)\gamma_5\tilde S_d(p) \rangle_{\rm amp}\,,
\end{split}
\end{gather}
where $\tilde S$ is the momentum-space quark propagator, and $S_{u-c}$ is a
shorthand for $S_u-S_c$. The appearance of the
integral over all momenta $q$ in the $(u-c)$ quark loop, appearing in the correlator
of $Q_1^{\pm}$, ensures that the propagator closes over itself. We will refer to the
two terms contributing to $F_1^{\pm}$ as ``connected'' and ``disconnected'', respectively.

Now we expand \req{eq:risubt_exp}
to order $g_{\rm s}^2$ in perturbation theory, with the notation
\begin{gather}
\chi = \chi^{(0)} + g_{\rm s}^2 \chi^{(1)} + \ldots
\end{gather}
for any quantity $\chi$. For convenience, the perturbative analysis
will be performed in Minkowski spacetime, and we will adopt the
conventions and QCD Feynman rules employed in~\cite{Peskin:1995ev} from now on.
All computations will be performed in Feynman gauge.
Using the fact that all renormalisation
constants are equal to unity at tree level, the order $g_{\rm s}^0$ term reads
\begin{gather}
F_1^{\pm;(0)} + \half c^{\pm;(0)}(m_u^2-m_c^2)
\left[(m_s+m_d)F_{\rm S}^{(0)}-(m_s-m_d)F_{\rm P}^{(0)}\right] = 0\,.
\end{gather}
It is trivial to check that\footnote{Note in passing that the correlator $F_{\rm P}$ is
identically zero due to parity conservation.}
\begin{gather}
F_1^{\pm;(0)} = 0\,,~~~~~~~~~~
F_{\rm S}^{(0)} = 4\,,~~~~~~~~~~
F_{\rm P}^{(0)} = 0\,,
\end{gather}
implying the (otherwise trivial) result $c^{\pm;(0)}=0$. Using the vanishing
of the mixing coefficient at tree level the $g_{\rm s}^2$ term simplifies
considerably, and one is left with
\begin{gather}
F_1^{\pm;(1)} \,+\, \half c^{\pm;(1)}(m_u^2-m_c^2)(m_s+m_d)F_{\rm S}^{(0)} = 0\,.
\end{gather}
One thus only has to determine the one-loop contributions to $F_1^{\pm}$.
Note that the form of the one-loop term is independent on whether the quark
masses in $Q_2^\pm$ are taken bare or renormalised --- i.e. the difference between
the two prescriptions is a two-loop effect. Recall also that subtraction coefficients
are expected to contain logarithmic divergences, and therefore $c^{\pm;(1)}$
should contain log terms that adjust the leading-order anomalous dimensions of the
subtracted four-fermion operators.\footnote{It is important to stress that the vertex function
in \req{eq:risubt} does not have to be finite; only physical amplitudes involving
renormalised subtracted operators need to.}

\begin{figure}[t!]
\begin{center}
\includegraphics[width=95mm]{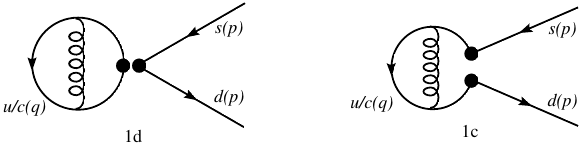}\vspace{5mm}
\includegraphics[width=95mm]{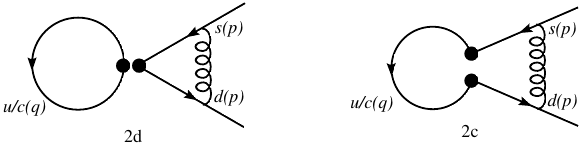}\vspace{5mm}
\includegraphics[width=95mm]{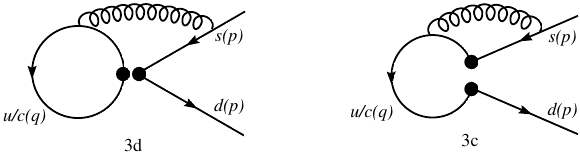}\vspace{5mm}
\includegraphics[width=95mm]{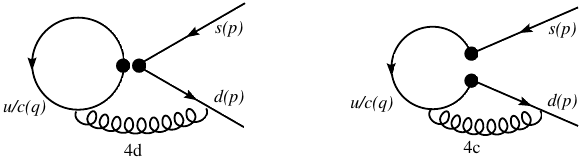}\vspace{5mm}
\includegraphics[width=95mm]{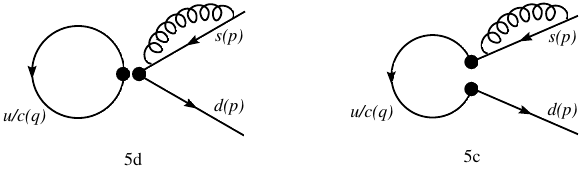}\vspace{5mm}
\includegraphics[width=95mm]{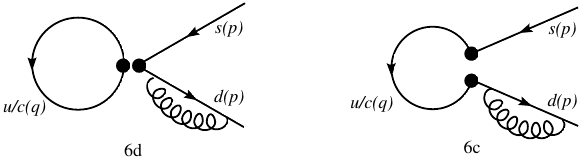}\vspace{5mm}
\end{center}
\vspace{-5mm}
\caption{One-loop diagrams contributing to $F_1^{\pm;(1)}$ (disconnected: left; connected: right; black dots signal insertions of $\gamma_\mu^{\rm L}$).}
\label{fig:oneloop_diag}
\end{figure}

The one-loop diagrams needed for the computation of $F_1^{\pm;(1)}$ are depicted
in~\refig{fig:oneloop_diag}. By writing the expression for each diagram
one immediately finds that diagrams 3d and 4d vanish because colour generators
at vertices lie in different colour traces; diagrams 1d and 1c vanish because
their spin traces are obviously zero; and diagrams 2d, 5d, 6d, 2c, 5c, and 6c
vanish because the expressions obtained are odd under $q\to -q$, and an integral
over $q$ is taken. One thus finds that the only contributions come from
diagrams 3c and 4c; denoting by $k$ the momentum carried by the gluon,
they read
\begin{align}
\nonumber
[{\rm 3c}] &= -{\rm tr}_{\rm c}[T_a T_a]\,\times \\
&~~~~~~
\int\kern-3pt\frac{\dif^4q}{(2\pi)^4}\int\kern-3pt\frac{\dif^4k}{(2\pi)^4}\,
\frac{{\rm tr}_{\rm s}[
\gamma^\nu(\slashed{p}-\slashed{k}+m_s)\gamma^{\mu\,{\rm L}}(\slashed{q}-\slashed{k}+m_{u/c})\gamma_\nu(\slashed{q}+m_{u/c})\gamma_\mu^{\rm L}
]}
{
D(k,0)D(p-k,m_s)D(q-k,m_{u/c})D(q,m_{u/c})
}\,,\\
\nonumber
[{\rm 4c}] &= -{\rm tr}_{\rm c}[T_a T_a]\,\times \\
&~~~~~~
\int\kern-3pt\frac{\dif^4q}{(2\pi)^4}\int\kern-3pt\frac{\dif^4k}{(2\pi)^4}\,
\frac{{\rm tr}_{\rm s}[
\gamma^{\mu\,{\rm L}}(\slashed{q}+m_{u/c})\gamma^\nu(\slashed{q}-\slashed{k}+m_{u/c})
\gamma_\mu^{\rm L}(\slashed{p}-\slashed{k}+m_d)\gamma_\nu
]}
{
D(k,0)D(p-k,m_d)D(q-k,m_{u/c})D(q,m_{u/c})
}\,,
\end{align}
where $T_a$ are the colour group generators normalised such that, for fundamental
quarks, ${\rm tr}_{\rm c}[T_a T_a]=(\NC^2-1)/2$; $D(l,m)=l^2-m^2+i\eta$;
and the result holds for either the $u$ or the $c$
quark circulating in the loop. After performing the Dirac traces and taking the
difference $(u-c)$, one ends up with
\begin{gather}
\begin{split}
F_1^{\pm;(1)} &= 8{\rm tr}_{\rm c}[T_a T_a]
\int\kern-3pt\frac{\dif^4q}{(2\pi)^4}\int\kern-3pt\frac{\dif^4k}{(2\pi)^4}\,
\frac{1}{D(k,0)}
\left(
\frac{m_s}{D(p-k,m_s)}\,+\,\frac{m_d}{D(p-k,m_d)}
\right)\,\\
&~~~~~~~~~~~~~~~~~~~~~~~~\times
\left(
\frac{m_u^2-q\cdot(q-k)}{D(q,m_u)D(q-k,m_u)}\,-\,
\frac{m_c^2-q\cdot(q-k)}{D(q,m_c)D(q-k,m_c)}
\right)\,.
\end{split}
\end{gather}
Note that, due to the vanishing of all disconnected contributions, the result
is the same for both operators $Q_1^\pm$.
Note also that both the $u$ and $c$ contributions separately lead to a quadratic
divergence, characteristic of the quark condensate, that explicitly cancels
after the difference $(u-c)$ is taken. Furthermore, some trivial algebra
allows to rewrite the two combinations containing the $(s,d)$ and $(u,c)$
contributions as
\begin{gather}
\begin{split}
\frac{m_s}{D(p-k,m_s)}\,&+\,\frac{m_d}{D(p-k,m_d)} =
\frac{m_s+m_d}{2}\,\times\\
&
\left\{
\frac{1}{D(p-k,m_s)}\,+\,\frac{1}{D(p-k,m_d)}\,+\,
\frac{(m_s-m_d)^2}{D(p-k,m_s)D(p-k,m_d)}
\right\}\,,
\end{split}
\end{gather}
and
\begin{gather}
\begin{split}
&\frac{m_u^2-q\cdot(q-k)}{D(q,m_u)D(q-k,m_u)}\,-\,
\frac{m_c^2-q\cdot(q-k)}{D(q,m_c)D(q-k,m_c)}
=
-\,\frac{m_u^2-m_c^2}{2}\,\times\\
&~~~~~~~~~~~~
\Bigg\{
\frac{1}{D(q,m_u)D(q,m_c)}\,+\,\frac{1}{D(q-k,m_u)D(q-k,m_c)}\\
&~~~~~~~~~~~~~~~~~~~~~~~~~~~~~
+\,\frac{k^2\left[m_u^2+m_c^2-k^2-2q^2+2(q\cdot k)\right]}{D(q,m_u)D(q,m_c)D(q-k,m_u)D(q-k,m_c)}
\Bigg\}\,,
\end{split}
\end{gather}
respectively. The expected mass dependence of the subtraction term thus
arises explicitly from the one-loop computation, and the final result
for the $\cO(g_{\rm s}^2)$ contribution to $c^\pm$ can be written as
\begin{gather}
c^{\pm;(1)}={\rm tr}_{\rm c}[T_a T_a]\left\{
2I_{\rm L}^{(2)}\left[I_{\rm L}^{(1)}+I_{\rm F}^{(1)}\right]
+I_{\rm F}^{(2)}
+I_{\rm F}^{(3)}
\right\}\,,
\end{gather}
with
\begin{align}
I_{\rm L}^{(1)} &= \int\kern-3pt\frac{\dif^4k}{(2\pi)^4}\,\frac{1}{D(k,0)}\left[\frac{1}{D(p-k,m_d)}+\frac{1}{D(p-k,m_s)}\right]\,,\\
I_{\rm L}^{(2)} &= \int\kern-3pt\frac{\dif^4q}{(2\pi)^4}\, \frac{1}{D(q,m_u)D(q,m_c)}\,,\\
I_{\rm F}^{(1)} &= \int\kern-3pt\frac{\dif^4k}{(2\pi)^4}\,
\frac{(m_s-m_d)^2}{D(k,0)D(p-k,m_d)D(p-k,m_s)}\,,\\
\nonumber
I_{\rm F}^{(2)} &= \int\kern-3pt\frac{\dif^4k}{(2\pi)^4}\int\kern-3pt\frac{\dif^4q}{(2\pi)^4}\,
\frac{-k^2+m_u^2+m_c^2-2q^2+2(q\cdot k)}{D(q,m_u)D(q,m_c)D(q-k,m_u)D(q-k,m_c)}\,\times\\
&\qquad\qquad\qquad\qquad\qquad
\left[
\frac{1}{D(p-k,m_d)}\,+\,\frac{1}{D(p-k,m_s)}
\right]\,,\\
\nonumber
I_{\rm F}^{(3)} &= \int\kern-3pt\frac{\dif^4k}{(2\pi)^4}\int\kern-3pt\frac{\dif^4q}{(2\pi)^4}\,
\frac{(m_s-m_d)^2}{D(p-k,m_d)D(p-k,m_s)}\,\times\\
&\qquad\qquad\qquad\qquad\qquad
\frac{\left[m_u^2+m_c^2-2q^2+2(q\cdot k)-k^2\right]}{D(q,m_u)D(q,m_c)D(q-k,m_u)D(q-k,m_c)}\,.
\end{align}
The integrals $I_{\rm F}^{(i)}$ are finite, while $I_{\rm L}^{(i)}$ are logarithmically
divergent.\footnote{The integral $I_{\rm F}^{(2)}$ seems to contain a divergent term
by power-counting in $k$, but it is easy to check that it is actually UV-finite.}
The latter can be worked out easily in dimensional regularisation;
for instance, taking the dimension over which the integral is performed as $D=4+2\epsilon$,
and denoting the subtraction point by $\mu$, one finds
\begin{align}
\nonumber
I_{\rm L}^{(1)} &= \frac{i\mu^{2\epsilon}}{(4\pi)^2}\,\bigg\{
-\,\frac{1}{\epsilon}-\gamma+\log(4\pi)\,+\\
&~~~~~~~~~~~~~~~~
+2-\log\left(\frac{p^2+m^2}{\mu^2}\right)
-\,\frac{m^2}{p^2}\,\log\left(1+\frac{p^2}{m^2}\right)
\bigg\}\,,\\
\nonumber
I_{\rm L}^{(2)} &= \frac{i\mu^{2\epsilon}}{(4\pi)^2}\,\bigg\{
-\,\frac{1}{\epsilon}-\gamma+\log(4\pi)\,+\\
&~~~~~~~~~~~~~~~~
+1-\frac{1}{m_u^2-m_c^2}\left[
m_c^2\log\left(\frac{m_c^2}{\mu^2}\right)-m_u^2\log\left(\frac{m_u^2}{\mu^2}\right)
\right]
\bigg\}\,,
\end{align}
where $\gamma \simeq 0.5772\ldots$ is the usual Euler-Mascheroni constant.

After reabsorbing the divergences consistently,
one is thus left with logarithm terms plus finite contributions.
By fixing $\mu^2=p^2$, it is easy to check that the log terms
in $I_{\rm L}^{(1)}$ vanish in the chiral limit, while those in
$I_{\rm L}^{(2)}$ contain an infrared divergence. This reflects the need of preserving
the flavour structure in the closed loop to avoid extra divergences from the quark
condensate. On the other hand, it is easy to check that there are no large logarithms.
Note also that finite contributions are suppressed
by an overall factor $p^{-2}$, and will become small for high
enough values of the external momentum. Finally, one crucial point is that there
are two loop integrals --- one from the gluon exchange and the other one over
the closed quark loop induced by the structure of the four-fermion operator.
Since each loop integral yields a factor $(4\pi)^{-2}$, this will make
the total one-loop correction $\propto g_{\rm s}^2/(4\pi)^4$ --- or, equivalently
\begin{gather}
c^{\pm;(1)} \sim \frac{\alpha_{\rm s}}{4\pi}\,\times\,\frac{\cO(1)}{(4\pi)^2}
\end{gather}
(no sign specified).
The extra factor of $(4\pi)^{-2}$ can be interpreted as a suppression of the one-loop
result with respect to its ``natural'' value $\alpha_{\rm s}/(4\pi)$.
This then supports the rough estimate that the subtraction coefficients
in the overlap regularisation, computed at hadronic scales, vanish up to an
$\alpha_{\rm s}/(4\pi)$ systematic uncertainty.

\section{Fits to ratios of correlation functions in the $p$-regime}
\label{app:fits}

In this appendix we provide some details of our fits to the ratios of correlation
functions $R_1^\pm$ involving four-quark operators, results for which are quoted in~\ret{tab:results_ratios},
for $p$-regime kinematics.\footnote{$\epsilon$-regime points are discussed in detail
in the main text.} Sufficiently far away from the insertions of kaon and
pion interpolating operators, such that all correlators are dominated by the
lowest-lying state in the corresponding channel, these ratios are expected
to become constant. To extract a value for the ratio of matrix elements,
we take an average over an interval in Euclidean time, using a jackknife
procedure to estimate errors that take statistical correlations into account properly.

Note that, since the contribution to $R_1^\pm$ from the eight-diagram provides a ratio of physical
amplitudes (it is proportional to the bag parameter for neutral meson oscillation), it will display
a plateau even if it is not combined with the contribution from the eye-diagram.
The latter will also display a plateau,
and it is possible to fit either contribution to a constant independently. This allows to better reconstruct
the contributions to the final noise-to-signal ratio in the quantities of interest.

\begin{figure}[t!]
  \begin{center}
    \begin{minipage}{.45\linewidth}
      \includegraphics[width=1.0\linewidth]{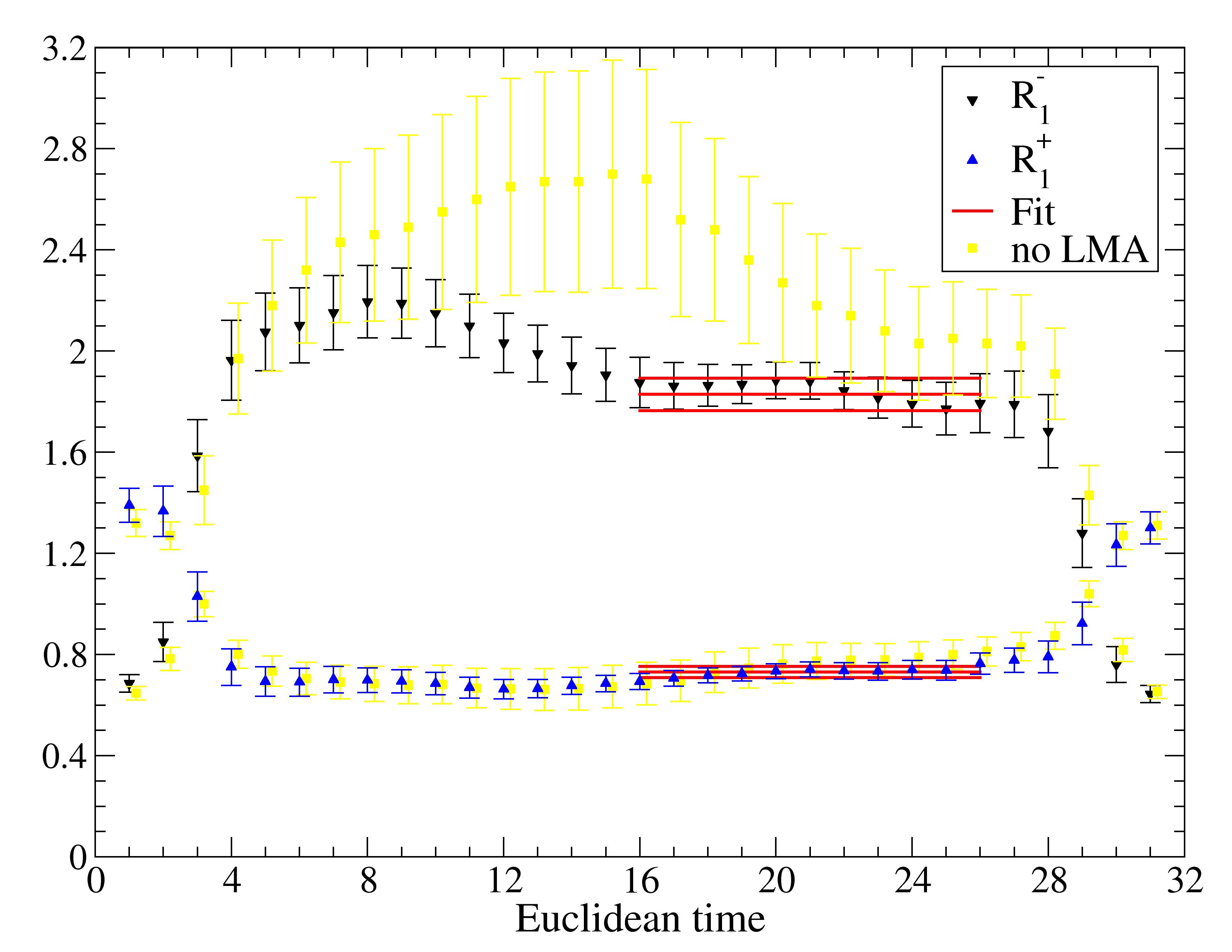}
	\end{minipage}
	\hspace{.05\linewidth}
    \begin{minipage}{.45\linewidth}
      \includegraphics[width=1.0\linewidth]{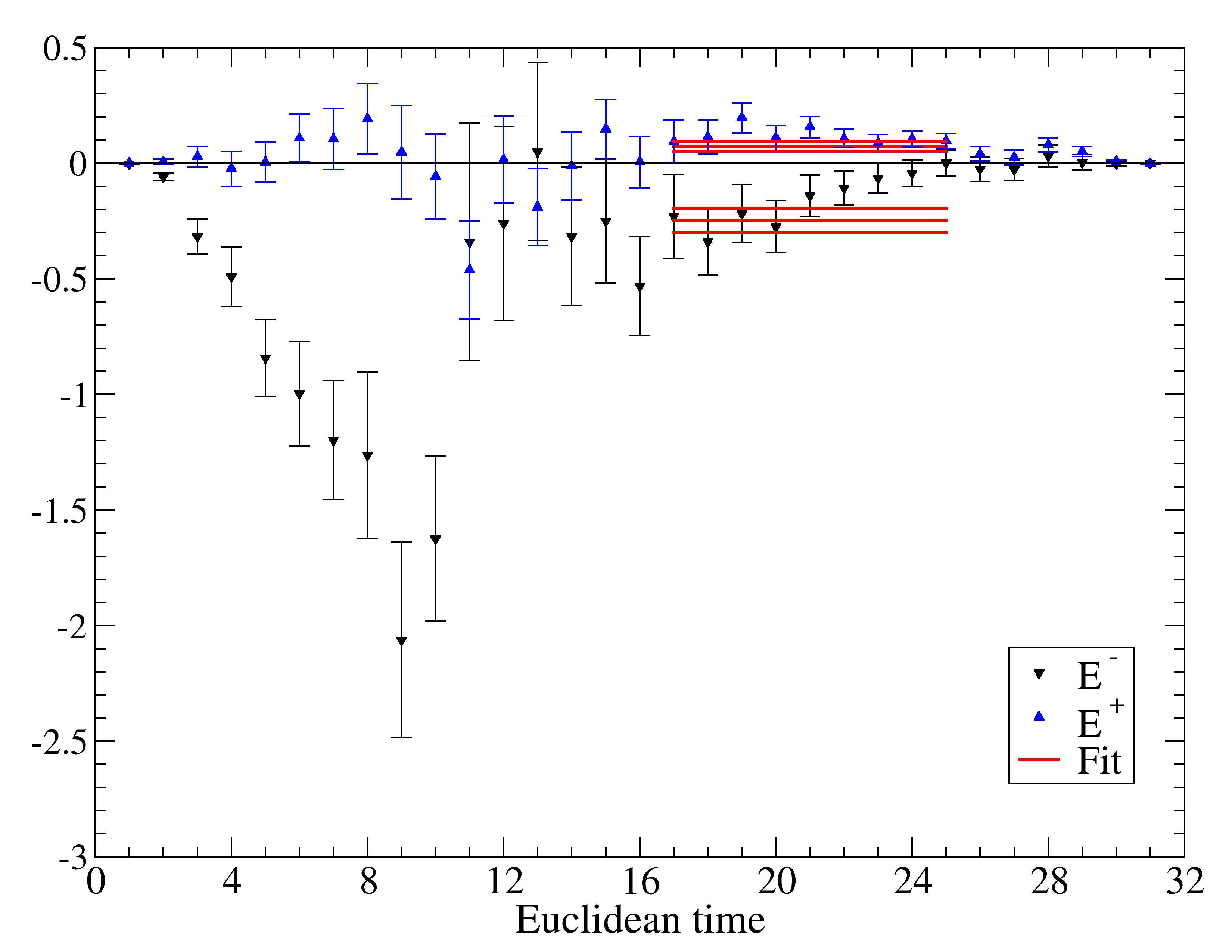}
	\end{minipage}
  \end{center}
\caption{Left: contribution to the ratios $R_1^\pm$ coming from the eight-diagram for $am_u=0.03$,
together with the fit to a plateau in some interval of Euclidean times. (The value of Euclidean
times is shifted by $10$ lattice units with respect to the conventions in the main text.)
The yellow points, corresponding to the computation that does not use low-mode averaging, illustrate
the impact of the latter on the signal.
Right: contribution to the ratios $R_1^\pm$ coming from the eye-diagram for $am_u=0.02, am_c=0.2$.}
\label{fig:plateaux}
\end{figure}

\begin{figure}[h!]
  \begin{center}
    \begin{minipage}{.45\linewidth}
      \includegraphics[width=1.0\linewidth]{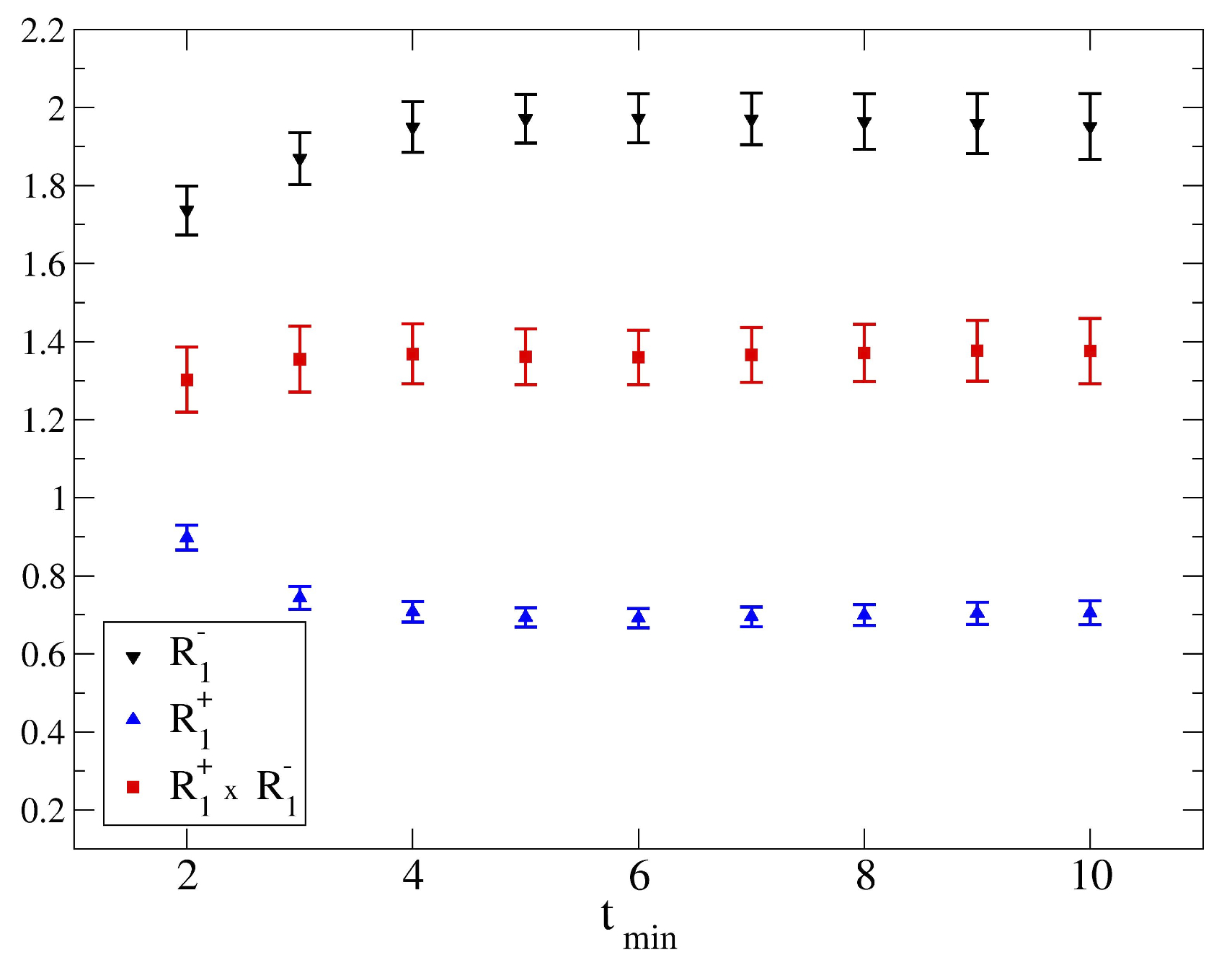}
	\end{minipage}
	\hspace{.05\linewidth}
    \begin{minipage}{.45\linewidth}
      \includegraphics[width=1.0\linewidth]{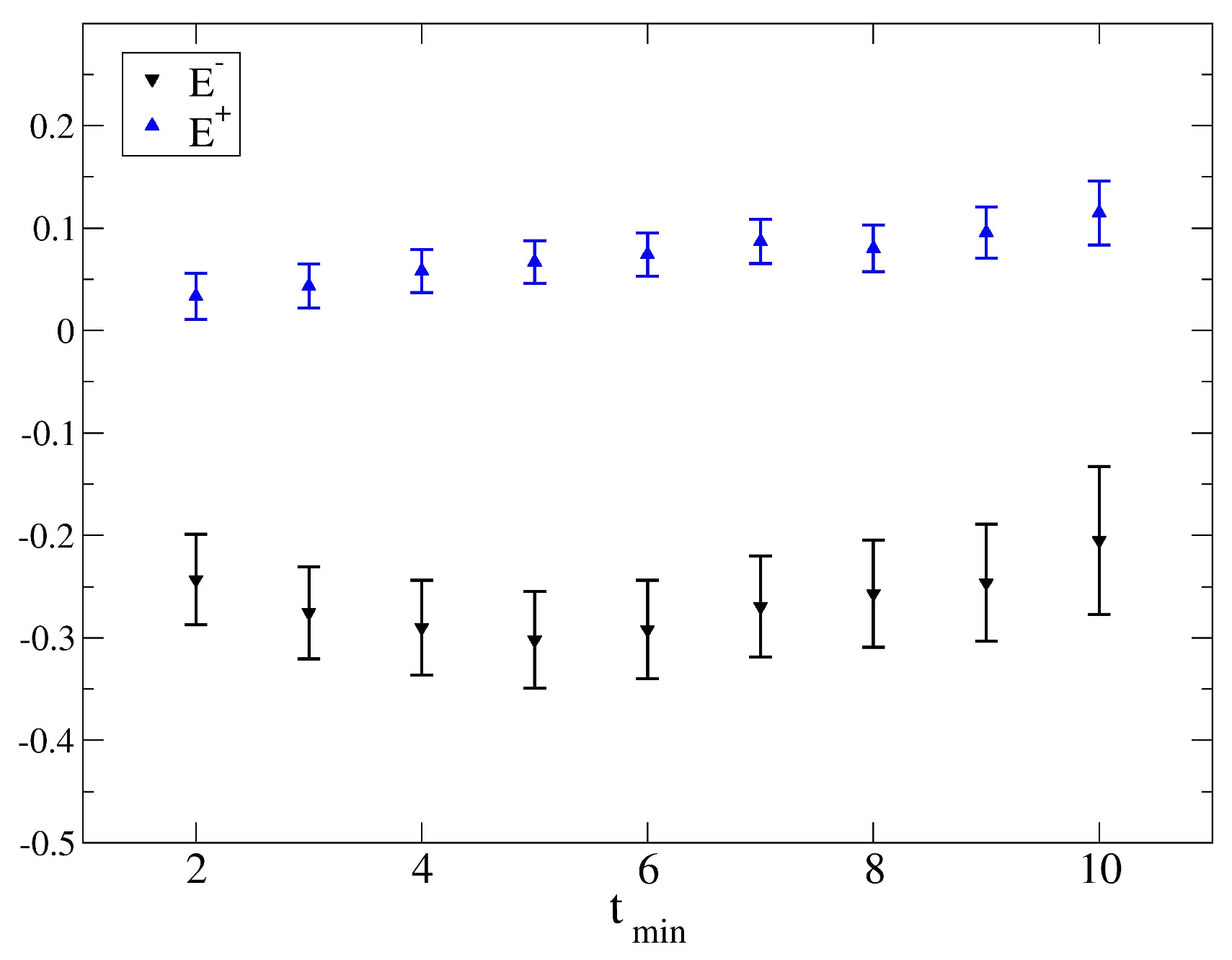}
	\end{minipage}
  \end{center}
\caption{Left: result for the fit to the contribution to $R_1^\pm$ ($am_u=0.03$) coming from the eight-diagram
as a function of the plateau choice, parametrised by the
minimal separation (in lattice units) $t_{\rm min}$ allowed between operator insertions.
The combination $R_1^++R_1^-$ is also displayed. Right:
idem for the eye-diagram ($am_u=0.03$, $am_c=0.2$).}
\label{fig:tdep}
\end{figure}

\refig{fig:plateaux} illustrates typical fits for both a numerically
well-behaved quantity (eight-diagrams for a not-too-light $p$-regime mass),
and a numerically challenging quantity (eye-diagrams with a large charm mass).
Note the sizeable errors, especially
in the case of contributions from the eye contraction. \refig{fig:tdep} shows
the dependence of the result on the choice of plateau, parametrised by the
minimal separation $t_{\rm min}$ (in lattice units) allowed between operator insertions.
Note that our LMA decomposition of correlation functions leads to some contributions
being known for all possible locations of the operator insertions; in those cases,
translational invariance has been exploited to improve the signal (although strong
correlations make the effect small). Given the very mild dependence of the results
on the choice of $t_{\rm min}$, provided the latter is large enough, we have chosen
the fit results for $t_{\rm sep} \in [6a, 10a]$ (in the notation used for correlation
functions in the main text) as representative, and quoted them
in~\ret{tab:results_ratios}. This is conservative, since taking a shorter interval
leads to the largest error and covers the systematic related to the plateau choice.

\end{appendix}
\bibliographystyle{JHEPjus}
\bibliography{biblio}
\end{document}